\documentclass[pra,preprint,floatfix,longbibliography]{revtex4}
\usepackage{hyperref} 
\usepackage{graphicx} 
\usepackage{subfig}
\usepackage{color}
\usepackage{filecontents}
\usepackage{soul}

\newcommand{\ket}[1]{\left| #1 \right\rangle}
\newcommand{\bra}[1]{\left\langle #1 \right|}

\newcommand{\be}{\begin{equation}}
\newcommand{\ee}{\end{equation}}
\newcommand{\bea}{\begin{eqnarray}}
\newcommand{\eea}{\end{eqnarray}}
\newcommand*{\myeqref}[2][Eq.~]{%
  \hyperref[{#2}]{#1(\ref*{#2})}%
}
\def\equationautorefname#1#2\null{%
  Eq.#1(#2\null)%
}

\usepackage{dcolumn}
\usepackage{bm}
\usepackage{amssymb,amsmath}
\usepackage{color}
\usepackage{float}
\usepackage{tikz}
\usetikzlibrary{arrows}

\definecolor{DarkGreen}{rgb}{0,0.6,0.2}

\begin{document}

\title{Influence of Disorder on Electromagnetically Induced Transparency in Chiral Waveguide Quantum Electrodynamics}
\author{Imran M. Mirza}
\affiliation{Department of Physics, University of Michigan, Ann Arbor, Michigan 48109, USA}
\email{imranmir@umich.edu}

\author{John C. Schotland}
\affiliation{Department of Mathematics and Department of Physics, University of Michigan, Ann Arbor, Michigan 48109, USA}
\email{schotland@umich.edu}
 
\begin{abstract}
We study single photon transport in a one-dimensional disordered lattice of three-level atoms coupled to an optical waveguide. In particular, we study atoms of $\Lambda$-type that are capable of exhibiting electromagnetically induced transparency (EIT) and separately consider disorder in the atomic positions and transition frequencies. We mainly address the question of how preferential emission into waveguide modes (chirality) can influence the formation of spatially localized states. Our work has relevance to experimental studies of cold atoms coupled to nano-scale waveguides and has possible applications to quantum communications.  
\end{abstract}

\maketitle

\section{Introduction}
Strong light-matter coupling plays a crucial role in quantum computing, quantum communication, and quantum information processing \cite{kimble2008quantum}. In the past, experimental setups based on cavity quantum electrodynamics (QED)~\cite{walther2006cavity} have been extensively used to realize this task~\cite{reiserer2015cavity} and strong coupling has been achieved even at the single photon level~\cite{kimble1998strong}. Recently, nanophotonic waveguides coupled to quantum emitters have emerged as a platform for quantum circuits~\cite{akimov2007generation,claudon2010highly,astafiev2010resonance,javadi2015single,goban2015superradiance}. Such so-called waveguide QED systems are also attractive due to the fact that they can support a continuum of optical modes and  can be used to construct quantum networks with applications to long-distance quantum communications~\cite{lodahl2015interfacing}.

In waveguide QED, light is strongly confined in the transverse plane of the waveguide and oscillates along the direction of propagation, due to enhancement of spin-orbit coupling~ \cite{lodahl2017chiral,petersen2014chiral, mitsch2014quantum, coles2016chirality}. In addition, in so-called chiral waveguides, light can propagate preferentially in one direction. There are multiple potential applications including devices that exhibit nearly unidirectional flow of light~\cite{sollner2015deterministic},  spontaneous and transient entanglement generators~\cite{gonzalez2015chiral, mirza2016multiqubit,mirza2016two}, and atom-photon circulators~\cite{scheucher2016quantum}.

To date, most work on multi-atom waveguide QED has concentrated on periodically arranged atoms in bidirectional waveguides with symmetric atom-waveguide coupling~\cite{roy2016strongly}. Many applications have been theoretically proposed and investigated experimentally. Examples include super and sub-radiance \cite{goban2015superradiance}, Bragg mirrors \cite{corzo2016large,sorensen2016coherent}, single photon transistors \cite {neumeier2013single,chang2007single} single-photon switches \cite{kim2010switching,liao2009quantum}, frequency comb generators \cite{liao2016single}, and single photon frequency converters \cite{bradford2012efficient}. In most instances, quantum emitters with two resonant or near-resonant energy levels are utilized as qubits. The presence of a third atomic level opens up new possibilities for quantum control and interference \cite{fang2016photon,martens2013photon,kolchin2011nonlinear,witthaut2010photon}. In particular, driven $\Lambda$-type atoms can manifest electromagnetically induced transparency (EIT)~\cite{fleischhauer2005electromagnetically}. The phenomenon of EIT is responsible for remarkable effects such as slow and stopped light \cite{hau1999light, kocharovskaya2001stopping}, enhanced optical nonlinearity \cite{jain1996efficient} and quantum memories \cite{lukin2003colloquium}.

For the case of periodically arranged atoms, a key aspect of the problem is to address the formation of allowed and forbidden bands. For symmetric waveguides coupled to three-level atoms, this problem has been discussed to some extent in the past \cite{witthaut2010photon}. However, the influence of chirality  on the band structure and dispersion has not been analyzed. In this work, we study this problem in detail and show that even a small chiral imbalance can introduce multiple resonances (with suppressed transmission), which is superimposed on the underlying EIT pattern. As the chiral imbalance is enhanced, the resonances form thin forbidden bands.

In this work, we study the problem of photon transport in disordered waveguide QED with $\Lambda$-type three-level atoms. Witthaut et al. have investigated this problem  in the setting of symmetric waveguides with atomic positional disorder\cite{witthaut2010photon}. They calculated the localization length as a function of frequency and have shown that localization can be controlled by an external drive. 
Here we analyze this problem in greater detail: we treat the possibility of disorder in atomic positions as well as in atomic transition frequencies. 
We calculate the localization length and single-photon transmission coefficient, both as a function of  photon frequency and the strength of the disorder. We also consider the effect of small back reflections and the effect of chirality on transport. Our results show both position and frequency disorder can significantly affect photon transport. In contrast, chiral waveguides are immune to the position disorder but show localization for frequency disorder. 

This paper is organized as follows. In Section II we introduce chiral waveguide and discuss photon transport in both periodic and disordered settings. In Section III, we address the bidirectional waveguide problem. In Section IV we focus on the band structure of periodic waveguides. We consider the effects of the disorder in Section V. Finally, in Section VI, we conclude with a summary of our results.

\section{Chiral waveguides}
\begin{figure*}
\centering
  \begin{tabular}{@{}cccc@{}}
   \includegraphics[width=6.5in, height=1.6in]{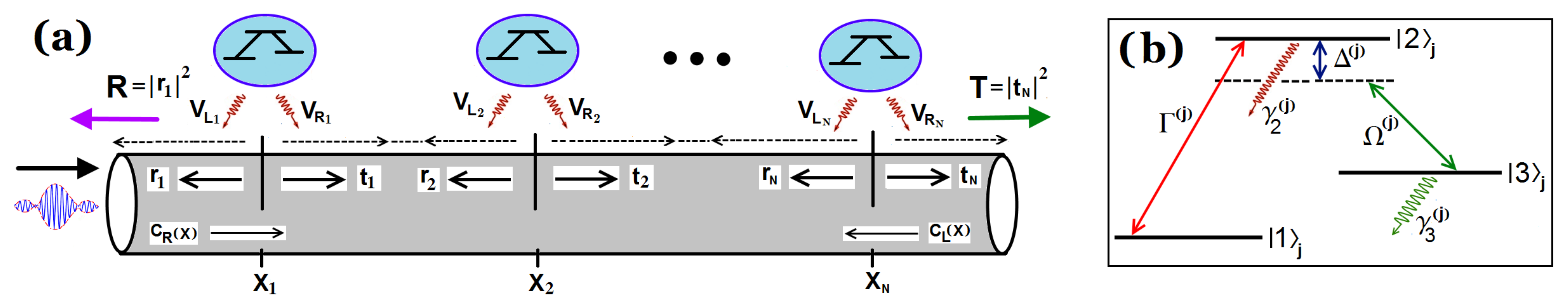} 
  \end{tabular}
\captionsetup{
  format=plain,
  margin=1em,
  justification=raggedright,
  singlelinecheck=false
}
 \caption{(Color online) (a) Illustrating the waveguide QED system considered in this paper. (b) Energy-level configuration of the $j$th three-level atom. }\label{Fig1}
\end{figure*}

In what follows, we first consider the problem of a chiral waveguide QED setup in which multiple three-level atoms are preferentially coupled to the waveguide without any back reflections (see Fig.\ref{Fig1}(a)). Note that such a condition is not far away from the experimental progress on chiral waveguide QED architectures. For instance, S\"ollner et al. have reported,  90\%  directionalities  and  98\%  atom-waveguide  coupling  strengths in  photonic  crystals  \cite{sollner2015deterministic}. \\
For a chiral waveguide the Hamiltonian in the real-space formalism \cite{shen2005coherentOL,shen2005coherent} and under the rotating wave approximation is given by
\begin{equation}
\begin{split}
&\hat{H}=\hbar\sum_{j}(\omega^{(j)}_{2}-i\gamma^{(j)}_{2})\hat{S}^{\dagger(j)}_{12}\hat{S}^{(j)}_{12}+\hbar\sum_{j}\omega^{(j)}_{3}\hat{S}^{\dagger(j)}_{23}\hat{S}^{(j)}_{23}+\hbar\sum_{j}\frac{\Omega_{j}}{2}(\hat{S}^{\dagger(j)}_{23}+\hat{S}^{(j)}_{23})\\
&+\hbar\int dx \hat{c}^{\dagger}(x)(\omega_{0}-iv_{g}\frac{\partial}{\partial x})\hat{c}(x)+\hbar\sum_{j}\int dx\delta(x-x_{j})\left(V_{j}\hat{c}^{\dagger}(x)\hat{S}^{(j)}_{12}+V^{\ast}_{j}\hat{S}^{\dagger}_{12}\hat{c}(x)\right),
\end{split}
\end{equation} 
where the transition from ground state $\ket{1}_{j}$ to excited state $\ket{2}_{j}$ is exhibited by the atomic lowering operator $\hat{S}^{(j)}_{12}\equiv\ket{1}_{j}\bra{2}$ (see Fig.\ref{Fig1}(b)). Whereas, the detuned transition from the excited state $\ket{2}_{j}$ to meta stable state $\ket{3}_{j}$ is derived from an external laser with Rabi frequency $\Omega^{(j)}$ (detuning $\Delta^{(j)}$) and transition operator $\hat{S}^{(j)}_{23}\equiv\ket{2}_{j}\bra{3}$. The decay rate from the excited (meta-stable) state is represented by $\gamma^{(j)}_{2}(\gamma^{(j)}_{3})$. The energy of the state $\ket{i}_{j}$ is taken to be $\hbar\omega^{(j)}_{i},(\forall i=1,2,3)$ where energy of the $\ket{1}_{j}$ is set to be zero. $v_{g}$ is the group velocity of the photon and $\omega_{0}$ is the frequency around which waveguide dispersion relation has been linearized. Destruction of photon in the waveguide continuum at position $x$ is represented by real-space annihilation operator $\hat{c}_(x)$. The $j$th emitter in the atomic chain is coupled to the waveguide continuum with an interaction strength $V_j$.

 
The field and atomic transition operators obey the following commutation relations:
\begin{equation}
\begin{split}
&[\hat{c}(x),\hat{c}^{\dagger}(x^{'})]=\delta(x-x^{'}),[\hat{S}^{\dagger(i)}_{12},\hat{S}^{(j)}_{12}]=\hat{S}^{(i)}_{12_{z}}\delta_{ij},\hspace{2mm}[\hat{S}^{\dagger(i)}_{23},\hat{S}^{(j)}_{23}]=\hat{S}^{(i)}_{23_{z}}\delta_{ij}.
\end{split}
\end{equation}

Note that the model of multiqubit waveguide QED shown in Fig.\ref{Fig1} can be realized in various physical systems for instance: Cesium atoms coupled to photonic crystal waveguide \cite{goban2015superradiance}, Cadmium Selenide quantum dots interacting with Ag nanowires \cite{akimov2007generation}, Artificial atoms (Josephson junctions) in microwave transmission lines \cite{astafiev2010resonance} and Silicon-
vacancy (SiV) color centers coupled to diamond nanodevices \cite{sipahigil2016integrated}. \\

The quantum state restricted to zero and one excitation in the system is given by
\begin{equation}
\begin{split}
&\ket{\Psi}=\int dx \varphi(x)\hat{c}^{\dagger}(x)\ket{\varnothing}+\sum_{j}a^{(j)}_{2}\hat{S}^{\dagger(j)}_{12}\ket{\varnothing}+\sum_{j}a^{(j)}_{3}\hat{S}^{\dagger(j)}_{13}\ket{\varnothing},
\end{split}
\end{equation}
where $\varphi(x)$, $a^{(j)}_{2}$ and $a^{(j)}_{3}$ are the amplitudes of finding a single excitation in the waveguide at the position $x$, the $j$th atom in the excited state $\ket{2}_{j}$, and $j$th atom in the metastable state $\ket{3}_{j}$. Note that $\hat{S}^{(j)}_{13}\equiv\ket{1}_{j}\bra{3}$ and $\ket{\varnothing}$ represents the ground state of the atom-waveguide system in which all atoms are unexcited and there are no photons in the waveguide. The equations obeyed by the above amplitudes are obtained from the time-independent Schr\"odinger equation $\hat{H}\ket{\Psi}=\hbar\omega\ket{\Psi}$ and are given by
\begin{subequations}
\label{eq:TIAE3}
\begin{eqnarray}
-iv_{g}\frac{\partial \varphi}{\partial x}+\sum_{j}V_{j}\delta(x-x_{j})a^{(j)}_{2}=(\omega-\omega_{0})\varphi(x),\label{eq:phic}\\
\frac{\Omega_{j}}{2}a^{(j)}_{3}+V^{\ast}_{j}\varphi(x=x_{j})=(\omega-\omega^{(j)}_{2}+i\gamma^{(j)}_{2})a^{(j)}_{2},\\
\frac{\Omega_{j}}{2}a^{(j)}_{2}=(\omega-\omega^{(j)}_{3})a^{(j)}_{3}.
\end{eqnarray}
\end{subequations}
Here $\omega$ is the frequency of the incoming photon. Next, we eliminate $a^{(j)}_{2}$ from Eq.~(\ref{eq:phic}) and obtain the equation followed by $\varphi$ as
{\begin{equation}
\label{eq:phiEq}
-iv_{g}\frac{\partial \varphi(x)}{\partial x}+\sum^{N}_{j=1}v_j\delta(x-x_{j}) \varphi(x)
=(\omega-\omega_{0})\varphi(x) ,
\end{equation}
where 
\begin{equation*}
v_j =\frac{4|V_{j}|^{2}\left(\omega-\omega^{(j)}_{3}\right)}{\left(\omega-\omega^{(j)}_{2}+i\gamma^{(j)}_{2}\right)\left(\omega-\omega^{(j)}_{3}\right)-\Omega^{2}_{j}}.
\end{equation*}
We obtain the solution to (\ref{eq:phiEq}) by observing that in between the atoms, when $x\neq x_j$,   $\varphi(x)=e^{iqx}$, where the wavenumber $q=(\omega-\omega_{0})/v_g$. Thus $\varphi$ takes the form
\begin{equation}
\label{eq:phi}
\varphi(x)=
\begin{cases}
  e^{iqx}, \hspace{5mm}x<x_1,\\      
  t_{1}e^{iqx}, \hspace{2mm} x_{1} \le x \le x_2, \\
  \vdots \\
  t_{N}e^{iqx}, \hspace{2mm}x>x_N  .
\end{cases}
\end{equation}
Next, we take $\epsilon$ as a small positive number and integrate (\ref{eq:phiEq}) over the interval $[x_j-\epsilon,x_j+\epsilon]$, to find the coefficients $t_j$. This gives the jump condition
\begin{equation}
\label{eq:boundaryC}
iv_{g}[\varphi(x_{j}+\epsilon)-\varphi(x_{j}-\epsilon)]=v_j\varphi(x_{j}) .
\end{equation}
Next, we regularize $\varphi$ as 
\begin{equation}
\varphi(x_{j})=\lim_{\epsilon\longrightarrow 0}\left[\varphi(x_{j}+\epsilon)+\varphi(x_{j}-\epsilon)\right]/2
\end{equation}
and introduce the quantity $\Gamma_{j}={|V_{j}|^{2}}/{2v_{g}}$. Eq.~(\ref{eq:boundaryC}) thus becomes
{\begin{equation}
\varphi(x_{j}+\epsilon)=T_j\varphi(x_{j}-\epsilon) ,
\end{equation}}
where 
\begin{equation}
\label{def_Tj}
T_j = \frac{\left(\omega-\omega^{(j)}_{2}+i\gamma^{(j)}_{2}\right)\left(\omega-\omega^{(j)}_{3}\right)-(\Omega_{j}/2)^{2}-i\left(\omega-\omega^{(j)}_{3}\right)\Gamma_{j}}{\left(\omega-\omega^{(j)}_{2}+i\gamma^{(j)}_{2}\right)\left(\omega-\omega^{(j)}_{3}\right)-(\Omega_{j}/2)^{2}+i\left(\omega-\omega^{(j)}_{3}\right)\Gamma_{j}} .
\end{equation} 
Then, using (\ref{eq:phi}) we arrive at the recursion relation
\begin{equation}
\label{eq:Ct&t}
t_{j}= T_j t_{j-1} ,
\end{equation}

Finally, we define the transmission coefficient $T=|\varphi(x_N)/\varphi(x_1)|^2$, which after making use of (\ref{eq:phi}) and (\ref{eq:Ct&t}) gives 
\begin{equation}
\label{eq:netTC}
T = \prod_{j=1}^N |T_j|^2 \ .
\end{equation}
Clearly, in the no loss situation i.e. when $\gamma^{(j)}_{2}=0$, the system acts as an all-pass filter. 
\subsection{Periodic arrangement}
We first consider a periodic chiral situation and discuss the single photon transmission properties. In Fig.~\ref{Fig2}(a) we plot transmission for a single atom. We notice that for the parameters of choice (as reported in \cite{witthaut2010photon}), the system shows EIT. Furthermore, based on how atom-waveguide coupling $\Gamma$ compares with the spontaneous emission rate $\gamma_{2}$, we define three regimes of transmission: (I) Under-coupled ($\gamma_{2}>\Gamma$) (II) Critically-coupled ($\gamma_{2}=\Gamma$) and (III) Over-coupled ($\Gamma>\gamma_{2}$). 
\begin{figure*}[t]
\centering
  \begin{tabular}{@{}cccc@{}}
   \includegraphics[width=3.25in, height=2.2in]{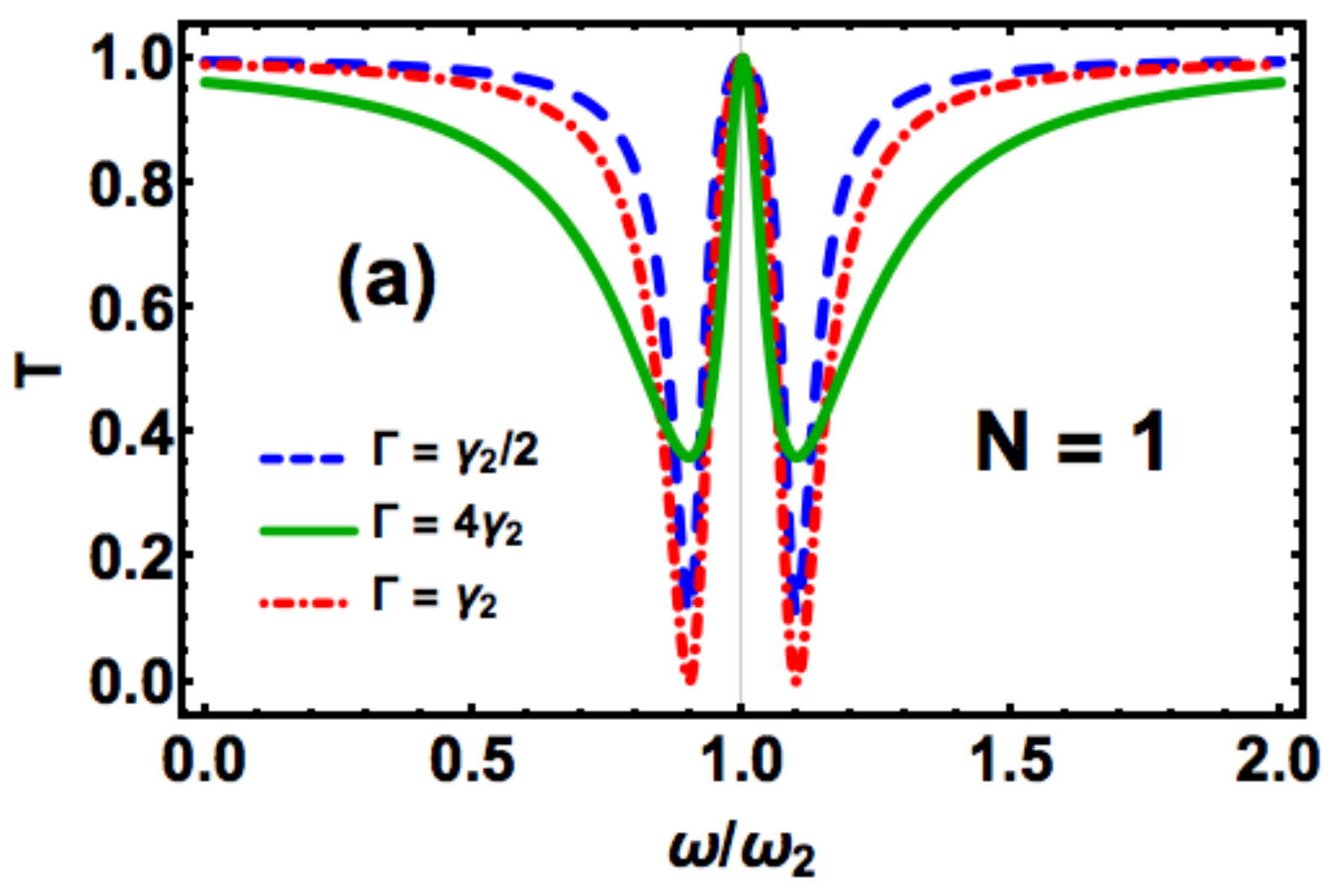} 
  \includegraphics[width=3.25in, height=2.2in]{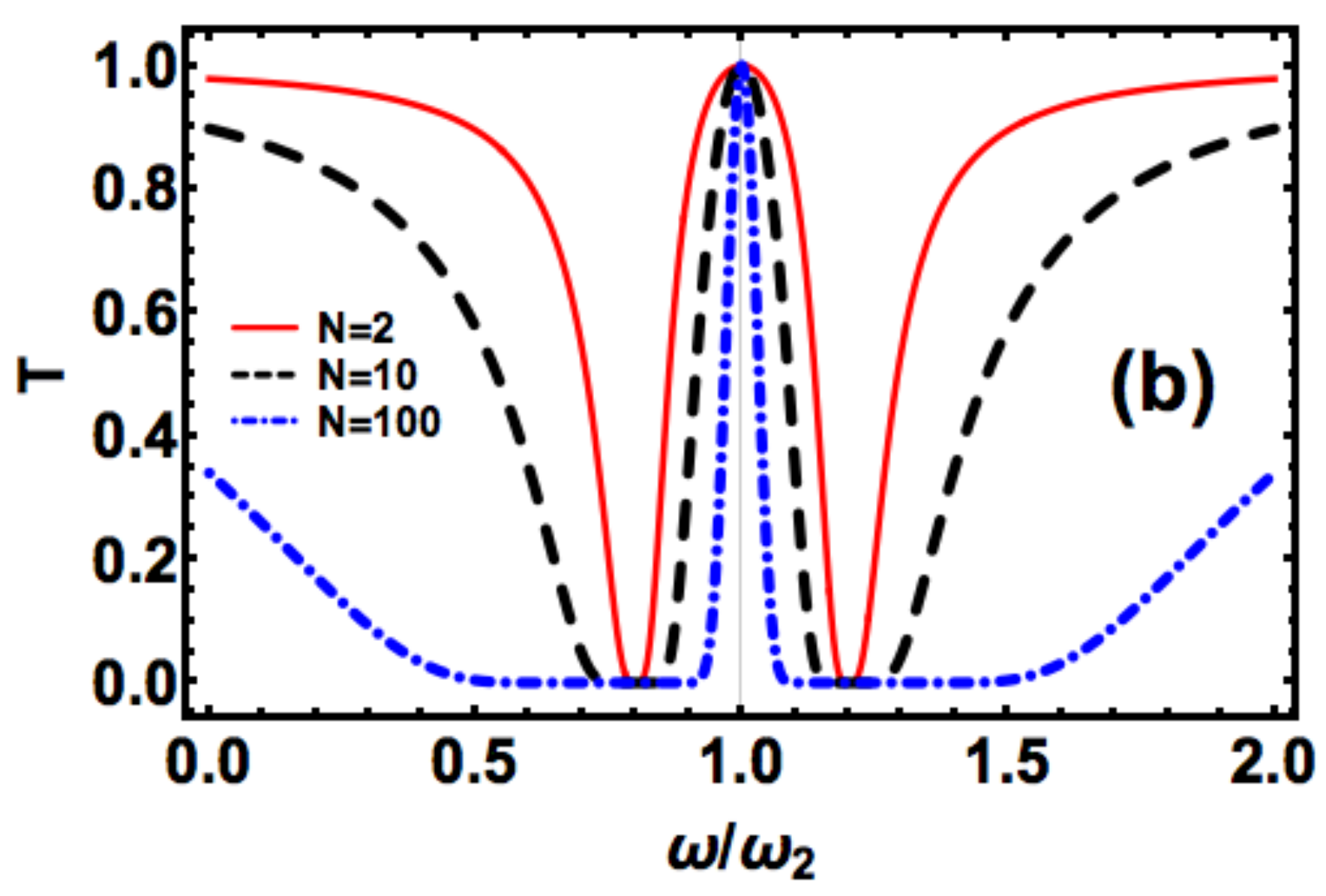}&
  \end{tabular}
\captionsetup{
  format=plain,
  margin=1em,
  justification=raggedright,
  singlelinecheck=false
}
 \caption{(Color online) Periodic chiral waveguide QED. (a) Transmission from a single atom. Green solid, blue dashed and red dotted dashed curves represent over ($\Gamma>\gamma_{2}$), under ($\Gamma<\gamma_{2}$) and critical ($\Gamma=\gamma_{2}$) coupling regimes, respectively.  The parameters are $\gamma_{2}=0.1\omega_{2}$ and $\Omega=0.2\omega_{2}$. (b) Transmission from a multiatom chain in the critical coupling regime with $\Omega=0.4\omega_{2}$ and $\gamma_{2}=0.1\omega_{2}$.}\label{Fig2}
\end{figure*}
Among these choices, transmission reaches the lowest value in the critical coupling regime at two $\omega_{2}$ points ($1.1\omega_{2}$ and $0.9\omega_{2}$) around the EIT peak. This behavior can be attributed to the complete destructive interference between the incoming field and the transmitted field at these two frequencies.\\
In Fig.~\ref{Fig2}(b) we plot the transmission for $N=2, 10$ and $100$ in the critical coupling regime. We observe as we increase the number of atoms, the width of the transparency window reduces while minimum transmission regions around the EIT point show growth.

\subsection{Disordered arrangement}
Next, we introduce disorder in the multi-atom chain and investigate single photon localization. For recent studies on the localization in photonic architectures, see for instance \cite{segev2013anderson,javadi2014statistical,lahini2008anderson,schwartz2007transport}. In what follows and for the rest of the paper, all random variables are generated from a Gaussian probability distribution of the form
\begin{equation}
\label{gaussian}
P(x)=\frac{1}{\sqrt{2\pi\sigma^2}}e^{-(x-\overline{x})^{2}/2\sigma^{2}},
\end{equation}
where $\sigma$ being the standard deviation is a measure of the strength of the disorder and $\overline{x}$ is the mean.
\subsubsection{Frequency disorder}
We start with the case of disorder in the atomic transition frequency $\omega_{2}$ (or equivalently in $\delta_{2}=\omega-\omega_{2}$). Such type of disorder can, for example, exist in optically trapped Rydberg's atom setups when trapping potential can be nonuniform, or when beam focusing is inhomogeneous \cite{zhang2011magic,maller2015rydberg}. Supposing that the detunings $\delta^{(j)}_{2}$ are independent and identically distributed random variables and using (\ref{def_Tj}) and (\ref{eq:Ct&t}), we find that the average transmission for an $N$-atom array can be expressed as 
\begin{eqnarray}
\langle T \rangle &=& \int \prod_{j=1}^N d\delta^{(j)}_2 P(\delta^{(j)}_2)|T_j|^2 \\
\label{TotalTC}&=& \langle |\tau|^2 \rangle^N.
\label{avgT}
\end{eqnarray}
where
\begin{equation}
\langle |\tau|^2 \rangle = \int d\delta_2 P(\delta_2) |\tau|^2  ,
\end{equation}
and 
\begin{equation}
\tau = \frac{\delta^{2}_2-(\Omega/2)^{2} + i (\gamma_2-\Gamma)\delta_2}{\delta^{2}_2-(\Omega/2)^{2} + i (\gamma_2+\Gamma)\delta_2} .
\end{equation}
We can easily write
\begin{equation}
|\tau|^{2}=1-\left[\left(\gamma_{2}+\Gamma\right)^{2}-\left(\gamma_{2}-\Gamma\right)^{2}\right]\delta^{2}_{2}\int^{\infty}_{0}e^{-\lambda\left((\delta^{2}_{2}-(\Omega/2)^{2})^{2}+(\gamma_{2}+\Gamma)^{2}\delta^{2}_{2}\right)}d\lambda .
\end{equation}
Performing the average over $\delta_{2}$, under the critical coupling and EIT condition with $\overline{\delta}_{2}=0$  gives
\begin{equation}
\begin{split}
&\langle |\tau|^2 \rangle =1-\frac{4\Gamma^{2}}{\sqrt{2\pi\sigma^{2}}}\int^{\infty}_{0}
\frac{{\rm exp}[\frac{1+8\lambda\Gamma^{2}\sigma^{2}-\lambda\sigma^{2}\Omega^{2}}{32\lambda\sigma^{4}}]}{32\lambda^{3/2}\sigma^{4}\sqrt{8\lambda\Gamma^{2}+\frac{1}{\sigma^{2}}-\lambda\Omega^{2}} } \Bigg[-(1+8\lambda\Gamma^{2}\sigma^{2}-\lambda\sigma^{2}\Omega^{2}) I_{-1/4}(z)\\
&+(1+16\lambda\sigma^{4}+64\lambda^{2}\Gamma^{4}\sigma^{4}-2\lambda\sigma^{2}\Omega^{2}+\lambda^{2}\sigma^{4}\Omega^{4}+16\lambda\Gamma^{2}\sigma^{2}-16\lambda^{2}\Gamma^{2}\Omega^{2}\sigma^{4})I_{1/4}(z)\\
&-(1+8\lambda\Gamma^{2}\sigma^{2}-\lambda\sigma^{2}\Omega^{2})\Bigg(I_{3/4}(z)+I_{5/4}(z)\Bigg)\Bigg]d\lambda ,
\end{split}
\end{equation}
where $I_{n}(z)$ is the modified Bessel function of the first kind with $z=\frac{(1+8\lambda\Gamma^{2}\sigma^{2}-\lambda\sigma^{2}\Omega^{2})^{2}}{32\lambda\sigma^{4}}$. Using (\ref{avgT}) the average transmission can be calculated.

As in the theory of disordered electronic system~\cite{markos2008wave,izrailev1999localization,delande2013many}, we define the localization length $\xi$ as
\begin{equation}
\label{llc}
\xi^{-1} = - \lim_{N\to\infty} \frac{\langle \ln T \rangle}{N} \ ,
\end{equation} 
where the average is performed over all detunings $\delta^{(j)}_{2}$. It is easily found that
\begin{equation}
\langle \ln T \rangle = N \langle \ln |\tau|^2 \rangle
\end{equation}
and hence
\begin{equation}
\xi^{-1} = - \langle \ln |\tau|^2 \rangle .
\end{equation}
In the critical coupling regime, we can perform the above average and thereby obtain
\begin{equation}
\begin{split}
&\xi^{-1}=-\frac{2\Gamma}{\sqrt{2\pi\sigma^2}}\int^{\infty}_{-\infty}\ln\left(1-\frac{4x^{2}}{(x^{2}-\Omega^{2}/16\Gamma^{2})^{2}+x^{2}}\right)e^{-(2\Gamma x-\overline{\delta_{2}})^{2}/2\sigma^{2}}dx.
\end{split}
\end{equation}
 
\begin{figure}[t]
\centering
  \begin{tabular}{@{}cccc@{}}
   \includegraphics[width=3.25in, height=2.2in]{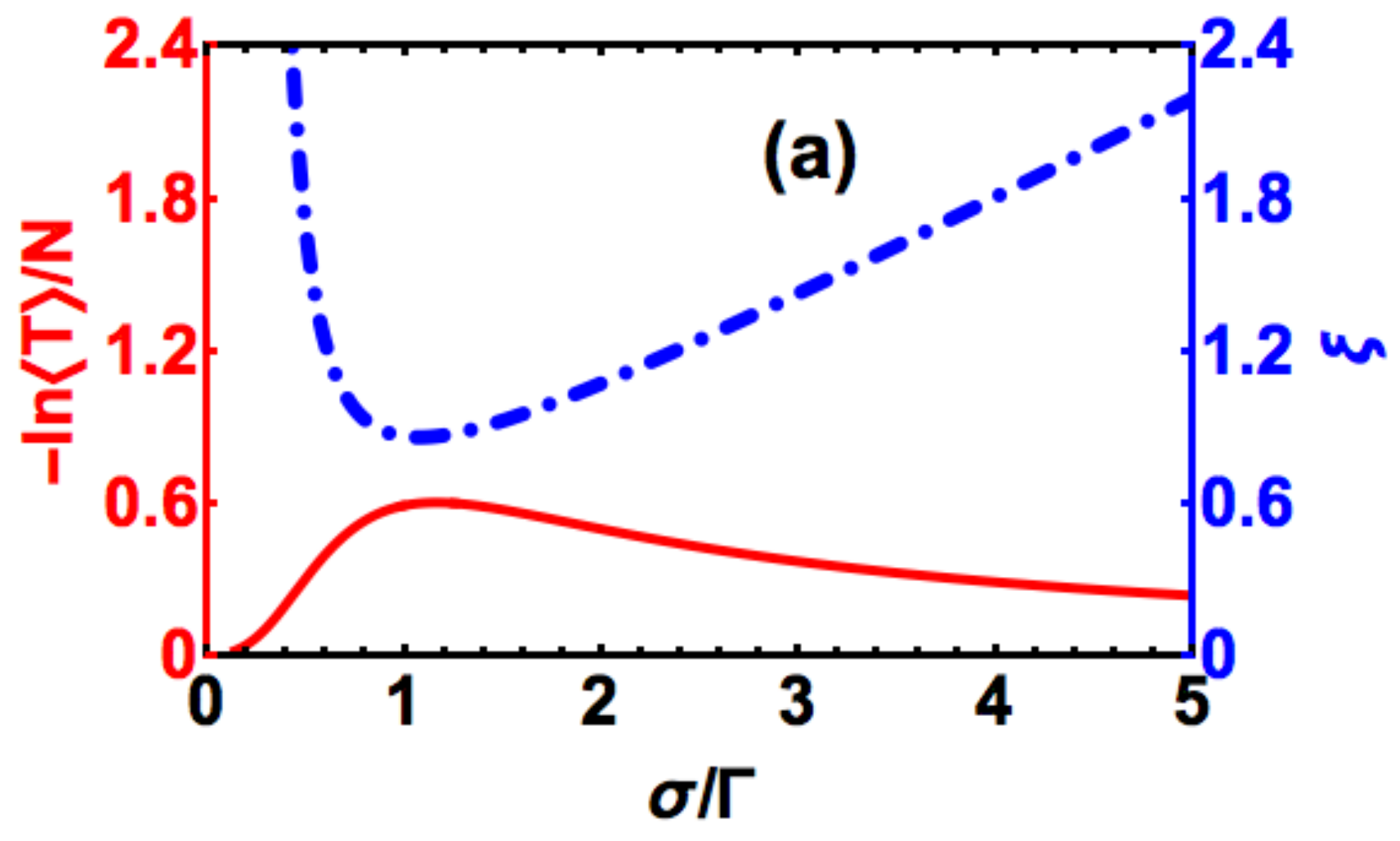} 
  \includegraphics[width=3.25in, height=2.2in]{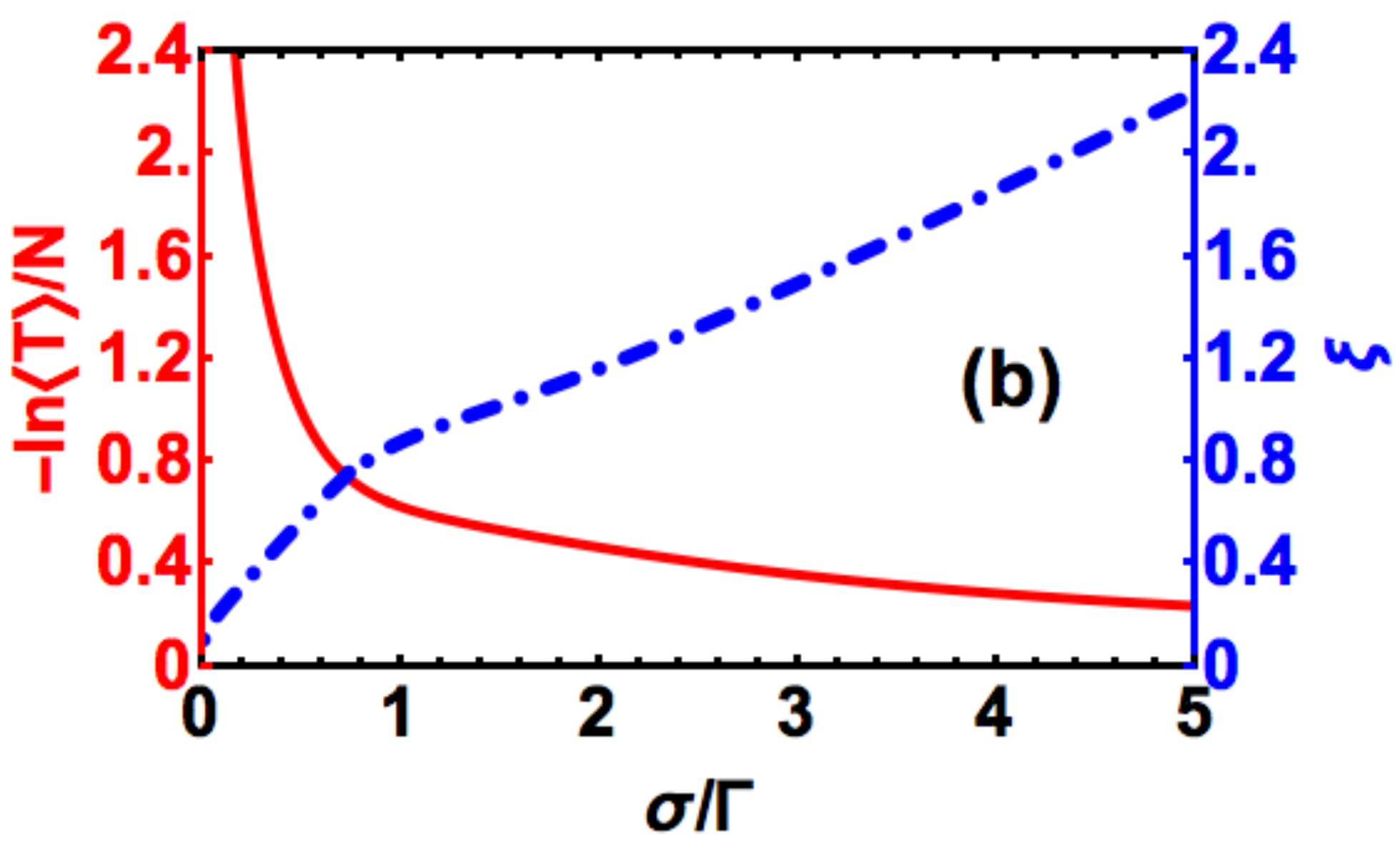}&
  \end{tabular}
\captionsetup{
  format=plain,
  margin=1em,
  justification=raggedright,
  singlelinecheck=false
}
 \caption{(Color online) Average transmission and localization length for a chiral waveguide QED with frequency disorder and EIT. In this plot the critical coupling regime has been considered with $\Omega=0.2\Gamma$ (a) $\overline{\delta_{2}}=0$ and (b) $\overline{\delta_{2}}=\Gamma$.}\label{Fig3}
\end{figure}
 In Fig.~\ref{Fig3}, we present plots of the average transmission and localization length as a function of disorder strength. When  $\overline{\delta_{2}}=0$, there is complete transmission for the perfectly ordered case. Since the conditions for EIT are satisfied, the transmission reaches its maximum. As the strength of the disorder is increased, the transmission decreases and correspondingly, the localization length also decreases. This result shows that frequency disorder can destroy  EIT in chiral waveguides.  In contrast, when $\overline{\delta_{2}}=\Gamma$, the average transmission starts at zero and grows as $\sigma$ is increased. 
 
\subsubsection{Position disorder}
Next, we consider the situation when the atomic positions are disordered. It follows from (\ref{eq:netTC}) that the transmission $T$ is independent of the atomic positions. Hence transmission in chiral waveguides is immune to position disorder. 

\section{Bidirectional waveguides}
We consider the following Hamiltonian for multiatom bidirectional waveguides
\begin{equation}
\label{eq:Hsys}
\begin{split}
\hat{H}=&\sum_{j}(\omega^{(j)}_{2}-i\gamma^{(j)}_{2})\hat{S}^{\dagger(j)}_{12}\hat{S}^{(j)}_{12}+\sum_{j}\omega^{(j)}_{3}\hat{S}^{\dagger(j)}_{23}\hat{S}^{(j)}_{23}+\sum_{j}\frac{\Omega_{j}}{2}(\hat{S}^{\dagger(j)}_{23}+\hat{S}^{(j)}_{23})\\
&+\int dx\hat{c}^{\dagger}_{R}(x)\left(\omega_{0} - iv_{R}\frac{\partial}{\partial x}\right)\hat{c}_{R}(x)  + \int dx\hat{c}^{\dagger}_{L}(x)\left(\omega_{0} + iv_{L}\frac{\partial}{\partial x}\right)\hat{c}_{L}(x)\\
&+\sum_{m,j}\int dx\delta(x-x_{j})\left[V_{mj}\hat{c}^{\dagger}_{m}(x)\hat{S}^{(j)}_{12}+h.c.\right] .
\end{split}
\end{equation}
The first two terms in (\ref{eq:Hsys}) represent the free Hamiltonian of the atoms. The third term accounts for the interaction of the external laser drive with the atoms. The fourth and fifth terms comprise the Hamiltonian of the waveguide, which supports left- and right-going modes with group velocities $v_R$ and $v_L$, respectively. Here  the sum is over $m\in\{R,L\}$. The annihilation of a single photon in the right (left) waveguide continuum at position $x$ is represented by the field operator $\hat{c}_{R}(x)(\hat{c}_{L}(x))$. The nonvanishing commutation relations for the field operators are 
\begin{eqnarray}
\left[\hat{c}_m(x),\hat{c}^\dag_{n}(x')\right] = \delta_{mn}\delta(x-x').
\end{eqnarray}
Finally, the last term in (\ref{eq:Hsys}) takes into account the atom- field interaction. Here $V_{mj}$ denotes the corresponding coupling, which we take to be real-valued. The waveguide described by the Hamiltonian (\ref{eq:Hsys}) is referred to as \emph{bidirectional}. If $v_R=v_L$ and $V_{Rj}=V_{Lj}$ the waveguide  is called \emph{symmetric}. Note that if $v_R$ or $v_L$  vanish, this corresponds to a chiral waveguide. 

The quantum state of a single photon is of the form
\begin{equation}
\label{eq:Psi(t)}
\begin{split}
&\ket{\Psi}=\sum_{m}\int dx \varphi_{m}(x)\hat{c}^{\dagger}_{m}(x)\ket{\varnothing}+\sum_{j}a^{(j)}_{2}\hat{S}^{\dagger(j)}_{12}\ket{\varnothing}+\sum_{j}a^{(j)}_{3}\hat{S}^{\dagger(j)}_{13}\ket{\varnothing}.
\end{split}
\end{equation} 
Here $\varphi_{R}(x),(\varphi_{L}(x))$ is the single photon amplitude in the right (left) waveguide continuum. Similar to section II, the the equations obeyed by the amplitudes are of the form
\begin{subequations}
\label{eq:TIAENC}
\begin{eqnarray}
-iv_{R}\frac{\partial \varphi_{R}(x)}{\partial x}+\sum_{j}V^{(j)}_{R}a^{(j)}_{2}\delta(x-x_{j})=(\omega-\omega_0)\varphi_{R}(x) , \\
iv_{L}\frac{\partial \varphi_{L}(x)}{\partial x}+\sum_{j}V^{(j)}_{L}a^{(j)}_{2}\delta(x-x_{j})=(\omega-\omega_0)\varphi_{L}(x) , \\ 
\frac{\Omega_{j}}{2}a^{(j)}_{3}+V^{(j)}_{R}\varphi_{R}(x_{j})+V^{(j)}_{L}\varphi_{L}(x_{j})=(\omega-\omega^{(j)}_{2}+i\gamma^{(j)}_{2})a^{(j)}_{2}, \\
\frac{\Omega_{j}}{2}a^{(j)}_{2}=(\omega-\omega^{(j)}_{3})a^{(j)}_{3}.
\end{eqnarray}
\end{subequations}
Eliminating $a^{(j)}_{2}$ from the above equations, we obtain the following equations obeyed by $\varphi_{R}$ and $\varphi_{L}$:
\begin{subequations}
\label{eq:TIAENC1}
\begin{eqnarray}
-iv_{R}\frac{\partial \varphi_{R}(x)}{\partial x}+\sum_{j}V^{(j)}_{R}\varpi^{(j)}\delta(x-x_{j})\left(V^{(j)}_{R}\varphi_{R}(x)+V^{(j)}_{L}\varphi_{L}(x)\right)=(\omega-\omega_0)\varphi_{R}(x), \hspace{10mm}\\
iv_{L}\frac{\partial \varphi_{L}(x)}{\partial x}+\sum_{j}V_{L_{j}}\varpi^{(j)}\delta(x-x_{j})\left(V^{(j)}_{R}\varphi_{R}(x)+V^{(j)}_{L}\varphi_{L}(x)\right)
=(\omega-\omega_0)\varphi_{L}(x), \hspace{10mm}
\end{eqnarray}
\end{subequations}
where 
\begin{equation*}
\varpi^{(j)}\equiv\frac{(\omega-\omega^{(j)}_{3})}{(\omega-\omega^{(j)}_{2}+i\gamma^{(j)}_{2})(\omega-\omega^{(j)}_{3})-(\Omega_{j}/2)^{2}}.
\end{equation*}
To solve the above equations, we note that between the atoms, when $x\neq x_{j}$, the amplitudes $\varphi_{R}(x)$ and $\varphi_{L}(x)$ are given by
$\varphi_{R}(x)=A_Re^{iq_{R}x}$ and $\varphi_{L}(x)=A_Le^{-iq_{L}x}$. The wavenumbers associated with the right and left field amplitudes are defined by $q_{R}=(\omega-\omega_{0})/v_{R}$, $q_{L}=(\omega-\omega_{0})/v_{L}$, respectively, where $A_R$ and $A_L$ are constants. Consequently, we find that
\begin{equation}
\label{eq:phiR}
\varphi_{R}(x)=
\begin{cases}
  e^{iq_{R}x}, \hspace{5mm}x<x_1,\\      
  t_{1}e^{iq_{R}x}, \hspace{2mm} x_{1} \le x \le x_2, \\
  \vdots \\
  t_{N}e^{iq_{R}x}, \hspace{2mm}x>x_N  .
\end{cases}
\end{equation}
and
\begin{equation}
\label{eq:phiL}
\varphi_{L}(x)=
\begin{cases}
  r_{1}e^{-iq_{L}x}, \hspace{5mm}x<x_1,\\      
  r_{2}e^{-iq_{L}x}, \hspace{2mm} x_{1} \le x \le x_2, \\
  \vdots \\
   r_{N}e^{-iq_{L}x}, \hspace{2mm} x_{N-1} \le x \le x_N, \\
  0, \hspace{2mm}x>x_N  .
\end{cases}
\end{equation}
where $r_{N+1}=0$ and $t_{0}=1$. See Fig.~\ref{Fig1}. To obtain the coefficients $t_{j}$ and $r_{j}$ we integrate (\ref{eq:TIAENC1}) over the interval $[x_{j}-\epsilon,x_{j}+\epsilon]$, which yields the following jump conditions
\begin{subequations}
\begin{eqnarray}
-iv_{R}\left[\varphi_{R}(x_{j}+\epsilon)-\varphi_{R}(x_{j}-\epsilon)\right]+V^{(j)}_{R}\varpi^{(j)}\left(V^{(j)}_{R}\varphi_{R}(x_{j})+V^{(j)}_{L}\varphi_{L}(x_{j}) \right)=0,\hspace{5mm}\\
iv_{L}\left[\varphi_{L}(x_{j}+\epsilon)-\varphi_{L}(x_{j}-\epsilon)\right]+V^{(j)}_{L}\varpi^{(j)}\left(V^{(j)}_{L}\varphi_{L}(x_{j})+V^{(j)}_{R}\varphi_{R}(x_{j}) \right)=0.\hspace{5mm}
\end{eqnarray}
\end{subequations}
Regularizing the discontinuity in $\varphi_{m}$ by 
\begin{equation}
\varphi_{m}(x)=\lim_{\epsilon\longrightarrow 0}\left[\varphi_{m}(x_{j}+\epsilon)+\varphi_{m}(x_{j}-\epsilon)\right]/2, 
\end{equation}
and introducing the quantities $\Gamma_{R_{j}}=V_{R_{j}}^{2}/2v_{R}$ and $\Gamma_{L_{j}}={V_{L_{j}}^{2}}/{2v_{L}}$, we find that
\begin{subequations}
\begin{eqnarray}
\varphi_{R}(x_{j}+\epsilon)=\left(\frac{1-i\Gamma^{(j)}_{R}\varpi^{(j)}}{1+i\Gamma^{(j)}_{R}\varpi^{(j)}} \right)\varphi_{R}(x_{j}-\epsilon)-i\sqrt{\frac{v_{L}}{v_{R}}}\frac{\sqrt{\Gamma_{R_{j}}\Gamma_{L_{j}}}\varpi^{(j)}}{1+i\Gamma^{(j)}_{R}\varpi^{(j)}}\left(\varphi_{L}(x_{j}+\epsilon)+\varphi_{L}(x_{j}-\epsilon)\right),\hspace{8mm}\\
\varphi_{L}(x_{j}+\epsilon)=\left(\frac{1+i\Gamma^{(j)}_{L}\varpi^{(j)}}{1-i\Gamma^{(j)}_{L}\varpi^{(j)}} \right)\varphi_{L}(x_{j}-\epsilon)+i\sqrt{\frac{v_{R}}{v_{L}}}\frac{\sqrt{\Gamma_{R_{j}}\Gamma_{L_{j}}}\varpi^{(j)}}{1-i\Gamma^{(j)}_{L}\varpi^{(j)}}\left(\varphi_{R}(x_{j}+\epsilon)+\varphi_{R}(x_{j}-\epsilon)\right) , \hspace{10mm}
\end{eqnarray}
\end{subequations}
Using Eqs.~(\ref{eq:phiR}) and (\ref{eq:phiL}) we obtain the recursion relations
\begin{subequations}
\label{eq:TIAENC2}
\begin{eqnarray}
t_{j}=\left(\frac{1-i\Gamma^{(j)}_{R}\varpi^{(j)}}{1+i\Gamma^{(j)}_{R}\varpi^{(j)}} \right)t_{j-1}-i\sqrt{\frac{v_{L}}{v_{R}}}\frac{\sqrt{\Gamma_{R_{j}}\Gamma_{L_{j}}}\varpi^{(j)}}{1+i\Gamma^{(j)}_{R}\varpi^{(j)}}\left(r_{j}e^{-i(q_{R}+q_{L})x_{j}}+r_{j+1}e^{-i(q_{R}+q_{L})x_{j}}\right),\\
r_{j+1}=\left(\frac{1+i\Gamma^{(j)}_{L}\varpi^{(j)}}{1-i\Gamma^{(j)}_{L}\varpi^{(j)}} \right)r_{j}+i\sqrt{\frac{v_{R}}{v_{L}}}\frac{\sqrt{\Gamma_{R_{j}}\Gamma_{L_{j}}}\varpi^{(j)}}{1-i\Gamma^{(j)}_{L}\varpi^{(j)}}\left(t_{j}e^{i(q_{R}+q_{L})x_{j}}+t_{j-1}e^{i(q_{R}+q_{L})x_{j}}\right).
\end{eqnarray}
\end{subequations}
Next, we write the transmission and reflection coefficients in terms of the phase accumulated by single photon while traveling through the waveguide between two consecutive atoms: 
\begin{equation}
t_{j}=\widetilde{t}_{j}e^{-i(q_{R}+q_{L})x_{j}/2},\hspace{5mm} r_{j}=\widetilde{r}_{j}e^{i(q_{R}+q_{L})x_{j-1}/2},
\end{equation}
which defines the quantities $\widetilde{r}_{j}$ and $\widetilde{t}_{j}$.
After some rearrangement, (\ref{eq:TIAENC2}) can be expressed in the form of the matrix recursion relation
\begin{equation}
\label{eq:TransEq}
\begin{pmatrix}
    \widetilde{t}_{j}     \\
    \widetilde{r}_{j+1}  \\
\end{pmatrix}=\mathcal{T}_{j}\begin{pmatrix}
\widetilde{t}_{j-1}\\
\widetilde{r}_{j}\\
\end{pmatrix}.
\end{equation} 
Here the transfer matrix $\mathcal{T}_{j}$ is given by
\begin{equation}
\label{eq:T1B}
\mathcal{T}_{j}=
\begin{pmatrix}
    m^{(j)}_{11}       & m^{(j)}_{12} \\
    m^{(j)}_{21}       & m^{(j)}_{22}  \\
\end{pmatrix}.
\end{equation}
Here the quantities
\begin{equation}
\alpha^{(j)}_{R/L} = \frac{1\mp i\Gamma^{(j)}_{R/L}\varpi^{(j)}}{1\pm i\Gamma^{(j)}_{R/L}\varpi^{(j)}}, \hspace{5mm}\beta^{(j)}_{R/L} = \frac{\sqrt{\Gamma^{(j)}_{R}\Gamma^{(j)}_{L}}\varpi^{(j)}}{1\pm i\Gamma^{(j)}_{R/L}\varpi^{(j)}}
\end{equation}
are needed to define the matrix elements
\begin{equation}
\begin{split}
&m^{(j)}_{11}=\left(\frac{\alpha^{(j)}_{R}+\beta^{(j)}_{R}\beta^{(j)}_{L}}{1-\beta^{(j)}_{R}\beta^{(j)}_{L}}\right)e^{i\phi_{j}},
\hspace{3mm}m^{(j)}_{12}=-i\sqrt{\frac{v_{L}}{v_{R}}}\left(\frac{\beta^{(j)}_{R}(1+\alpha^{(j)}_{L})}{1-\beta^{(j)}_{R}\beta^{(j)}_{L}}\right)e^{-i\phi_{j}},\\
&m^{(j)}_{21}=i\sqrt{\frac{v_{R}}{v_{L}}}\left(\frac{\beta^{(j)}_{L}(1-\alpha^{(j)}_{R})}{1+\beta^{(j)}_{L}\beta^{(j)}_{R}}\right)e^{i\phi_{j}},\hspace{3mm}m^{(j)}_{22}=\left(\frac{\alpha^{(j)}_{L}-\beta^{(j)}_{L}\beta^{(j)}_{R}}{1+\beta^{(j)}_{L}\beta^{(j)}_{R}}\right)e^{-i\phi_{j}},\\
&\phi_{j}=(q_{R}+q_{L})(x_{j}-x_{j-1})/2 .
\end{split}
\end{equation}
Note that for symmteric waveguides, when $v_{L}=v_{R}$ (or equivalnetly $\Gamma_{R_{j}}=\Gamma_{L_{j}}$), we recover the results of Witthaut et al. \cite{witthaut2010photon}. The net transfer matrix $M$ for an $N$ atom system is given by
\begin{equation}
M=\prod_{j}\mathcal{T}_j :=
\begin{pmatrix}
   M_{11}       & M_{12} \\
    M_{21}      & M_{22}  \\
\end{pmatrix}.
\end{equation}
The net transmission coefficient is given by $T=|t_{N}|^{2}$ while the reflection coefficient is $R=|r_{1}|^{2}$. Note that $0\leq T\leq 1$. Note also that in the absence of  spontaeous emission, when $\gamma_{2}=0$, it can be seen that $T=\vert 1/M_{22}\vert^{2}$, which is a general property of transfer matrices \cite{markos2008wave}.

\section{Band Structure}
In this section, we consider a periodic arrangement of atoms and study the band structure in bidirectional waveguides. We first direct our attention to the single photon dispersion properties and then consider the influence of back reflections. For a discussion of dispersion properties of single photons in two- and three-level atoms coupled to symmetric waveguides,  we direct the reader to \cite{shen2005coherent, yanik2004stopping,zueco2012microwave,witthaut2010photon}.

\subsection{Dispersion relation}
To study the dispersion properties of a single photon, we make use of the periodicity of the lattice and consider solutions of the form
\begin{equation}
\widetilde{t}_{j}=te^{ijKL},  \quad  \widetilde{r}_{j+1}=re^{ijKL},
\end{equation}
where $K$ is the wavenumber and $L$ is the lattice spacing. By inserting these solutions in (\ref{eq:TransEq}), we find
\begin{equation}
\label{eq:disp2}
\begin{pmatrix}
    t   \\
    r     \\
\end{pmatrix}=e^{-i{K}L}\mathcal{T}
\begin{pmatrix}
    t   \\
    r     \\
\end{pmatrix},
\end{equation}
which shows that $e^{i{K}L}$ is an eigenvalue of $\mathcal{T}$. Thus
\begin{equation}
\label{detDR}
\det(\mathcal{T}-e^{i{K}L}\mathcal{I})=0,
\end{equation}
where $\mathcal{I}$ is the $2\times 2$ identity matrix. Eq.~(\ref{detDR}) thus becomes, for $\gamma_{2}=0$, the dispersion relation 
\begin{equation}
\cos({K}L)=\frac{\lbrace\Lambda^{4}-(\Gamma^{2}_{R}-\Gamma^{2}_{L})(\delta_{2}/4)^{2}\rbrace {\rm cos}(q_{R}+q_{L})L/2 + \lbrace\Lambda^{2}\Gamma_{R}\delta_{2}/2\rbrace {\rm sin}(q_{R}+q_{L})L/2}{ \Lambda^{4}+(\Gamma_{R}-\Gamma_{L})^{2}(\delta_{2}/4)^{2}    },
\end{equation}
where $\Lambda^{2}\equiv\delta^{2}_{2}-(\Omega/2)^{2}$. This result agrees with the dispersion relation for symmetric waveguides ($\Gamma_{R}=\Gamma_{L}$) as reported in \cite{witthaut2010photon}.

\subsection{Small back reflections}
In what follows, we consider a bi-directional waveguide subject to small back reflections ($\Gamma_{L}<\Gamma_{R}$). In Fig.~\ref{Fig4}(a), (b) and (c) we plot the frequency dependence of the  transmission and reflection coefficients as a function of the number of atoms. For a single emitter, we find that the location of the EIT peak remains unaffected by small back reflections. Additionally, the system shows $\sim 10$\% transmission at the two minima. Next, in Fig.~\ref{Fig4}(b) and (c) we focus on the multi-emitter problem and observe the formation of small resonances superimposed on the EIT pattern. For $N=50$, the EIT peak becomes sharper and the resonances where the transmission is suppressed become more visible.
\begin{figure*}
\centering
  \begin{tabular}{@{}cccc@{}}
   \includegraphics[width=1.6in, height=1.5in]{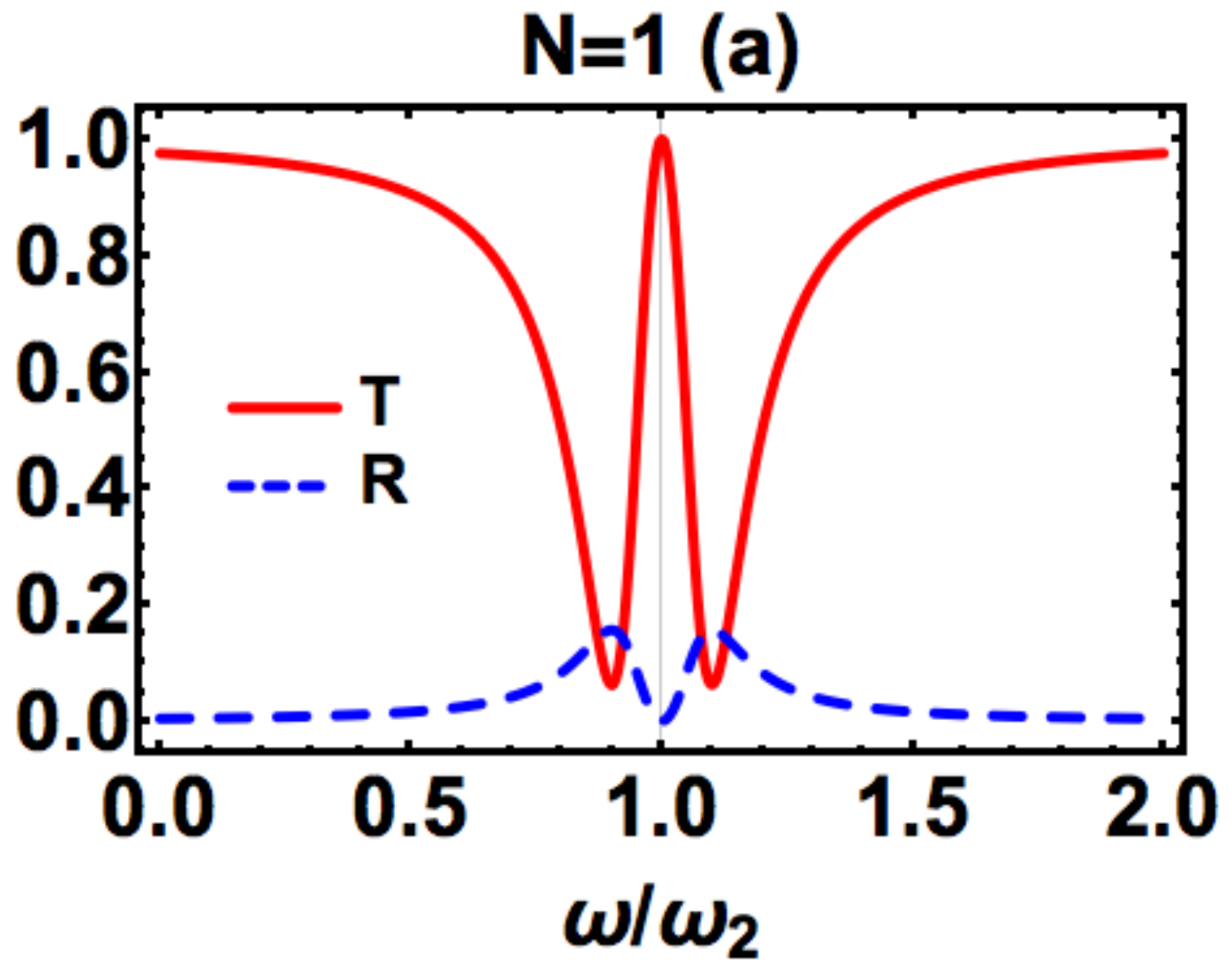} 
  \includegraphics[width=1.6in, height=1.5in]{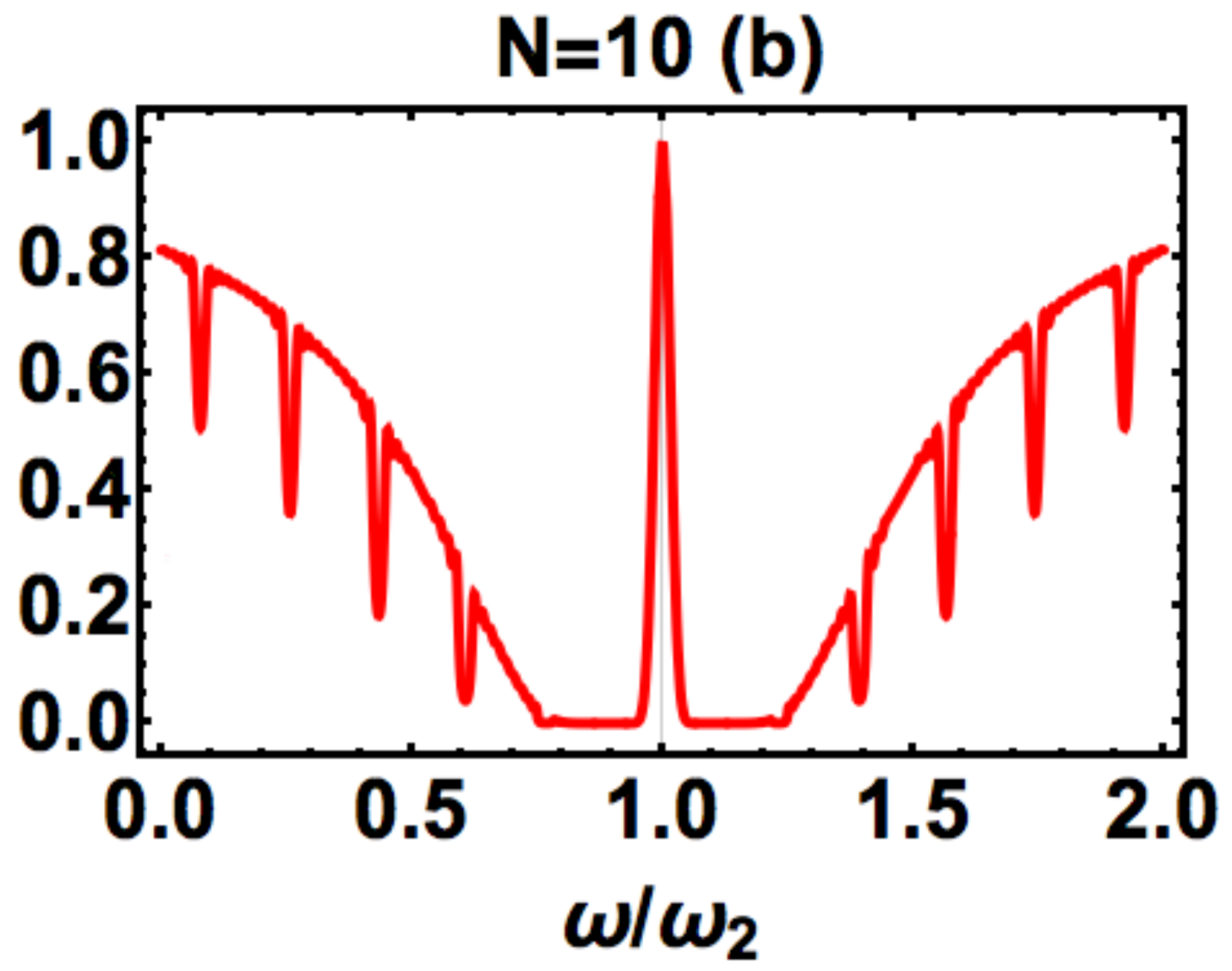}&
   \includegraphics[width=1.6in, height=1.5in]{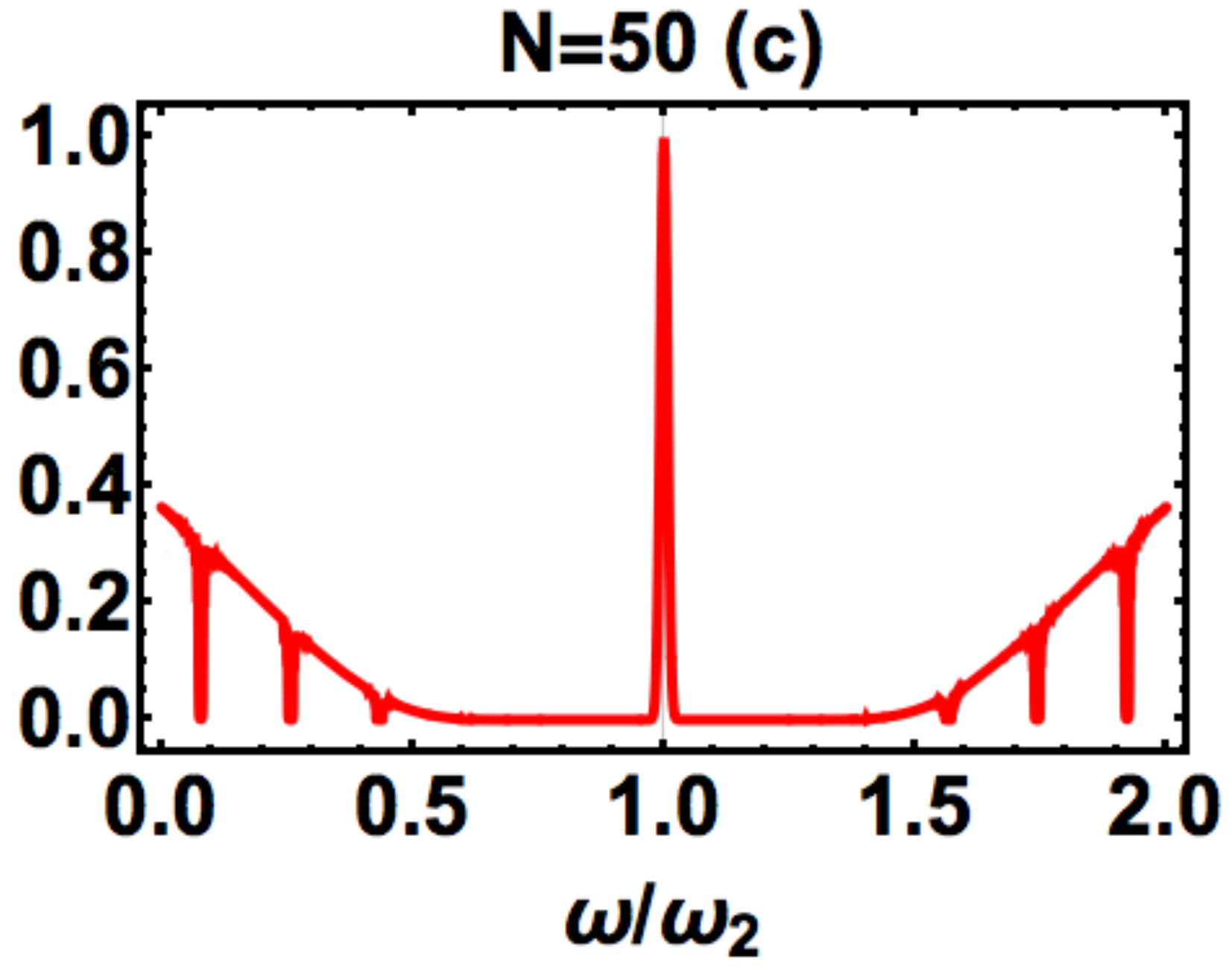}&
   \includegraphics[width=1.45in, height=1.8in]{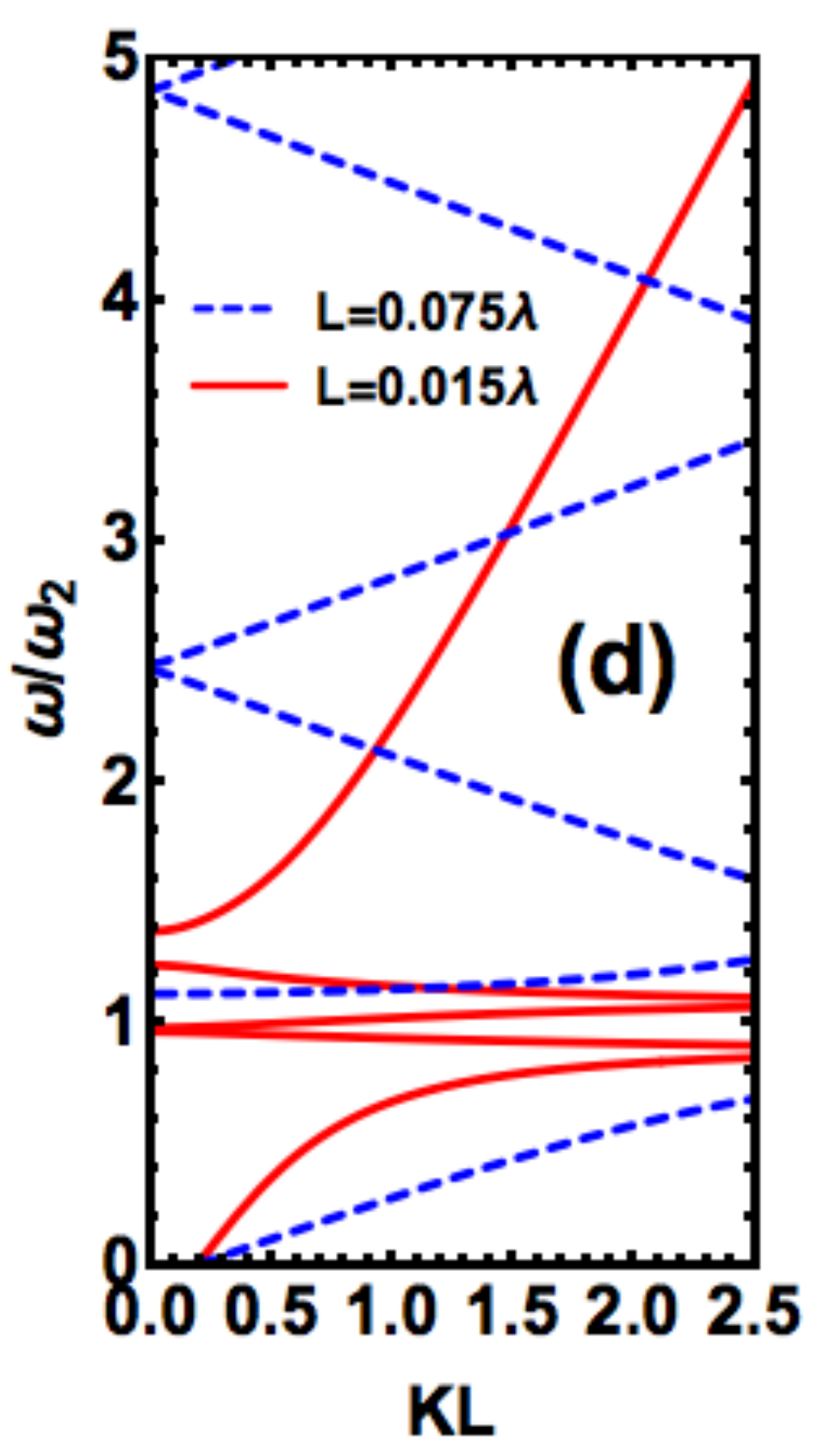}&
  \end{tabular}
\captionsetup{
  format=plain,
  margin=1em,
  justification=raggedright,
  singlelinecheck=false
}
 \caption{(Color online) Single-photon reflection and transmission in a small back reflecting waveguide with (a) $N=1$, (b) $N=10$ and (c) $N=50$ periodically arranged identical emitters. The following parameters are used:
  $\gamma_{2}=0.1\omega_{2}$, $\Gamma_{R}=0.4\omega_{2}$, $\Omega=0.2\omega_{2}$, $v_{L}=10v_{R}$ (or equivalently $\Gamma_{L}=0.1\Gamma_{R}$) and lattice constant $L=0.5\lambda$ (while $\lambda\equiv 2\pi v_{R}/\omega_{2}$). (d) Dispersion profile for two different inter-atomic distances.}
   \label{Fig4}
\end{figure*}
\begin{figure*}[h]
\centering
  \begin{tabular}{@{}cccc@{}}
   \includegraphics[width=1.6in, height=1.5in]{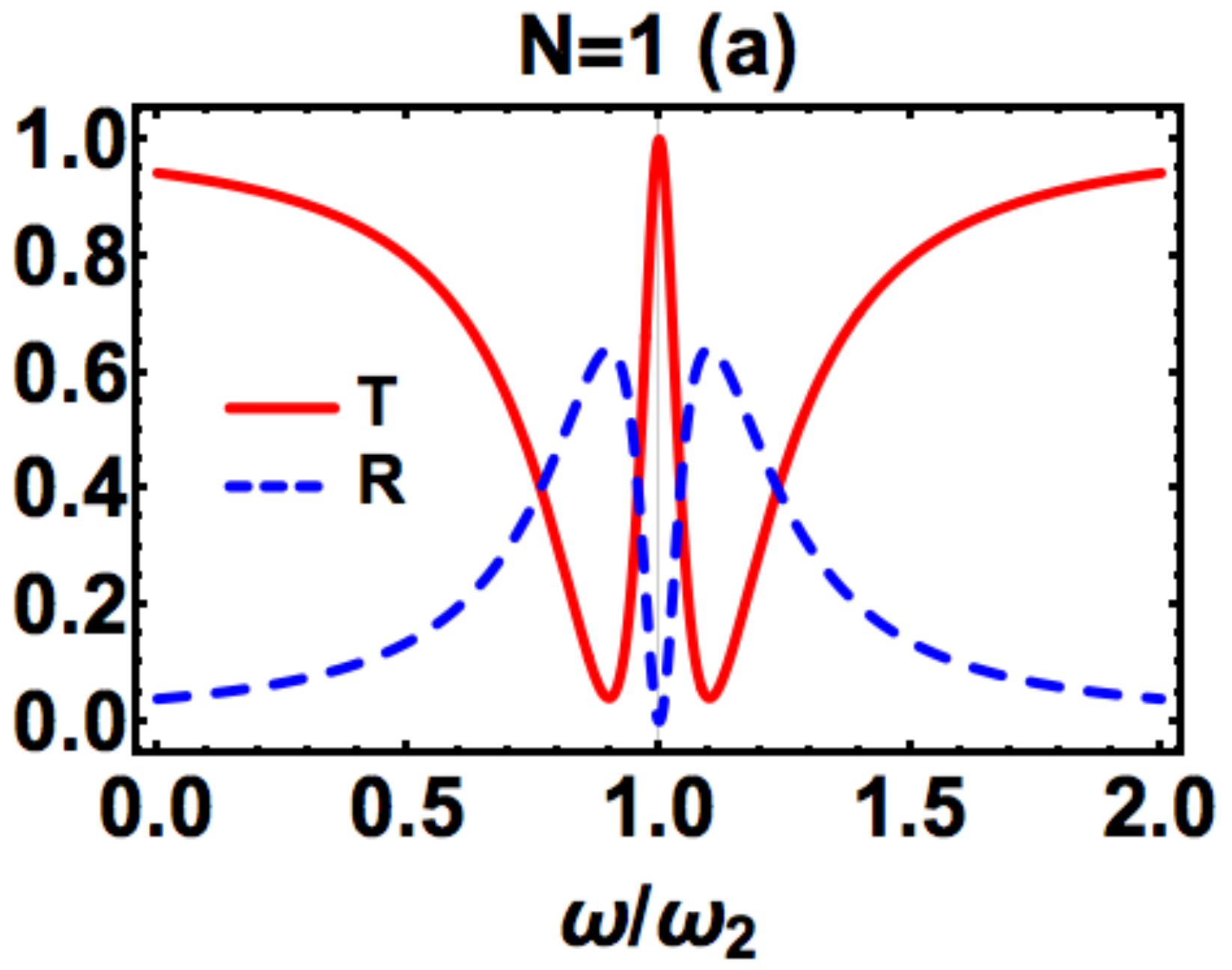} 
  \includegraphics[width=1.6in, height=1.5in]{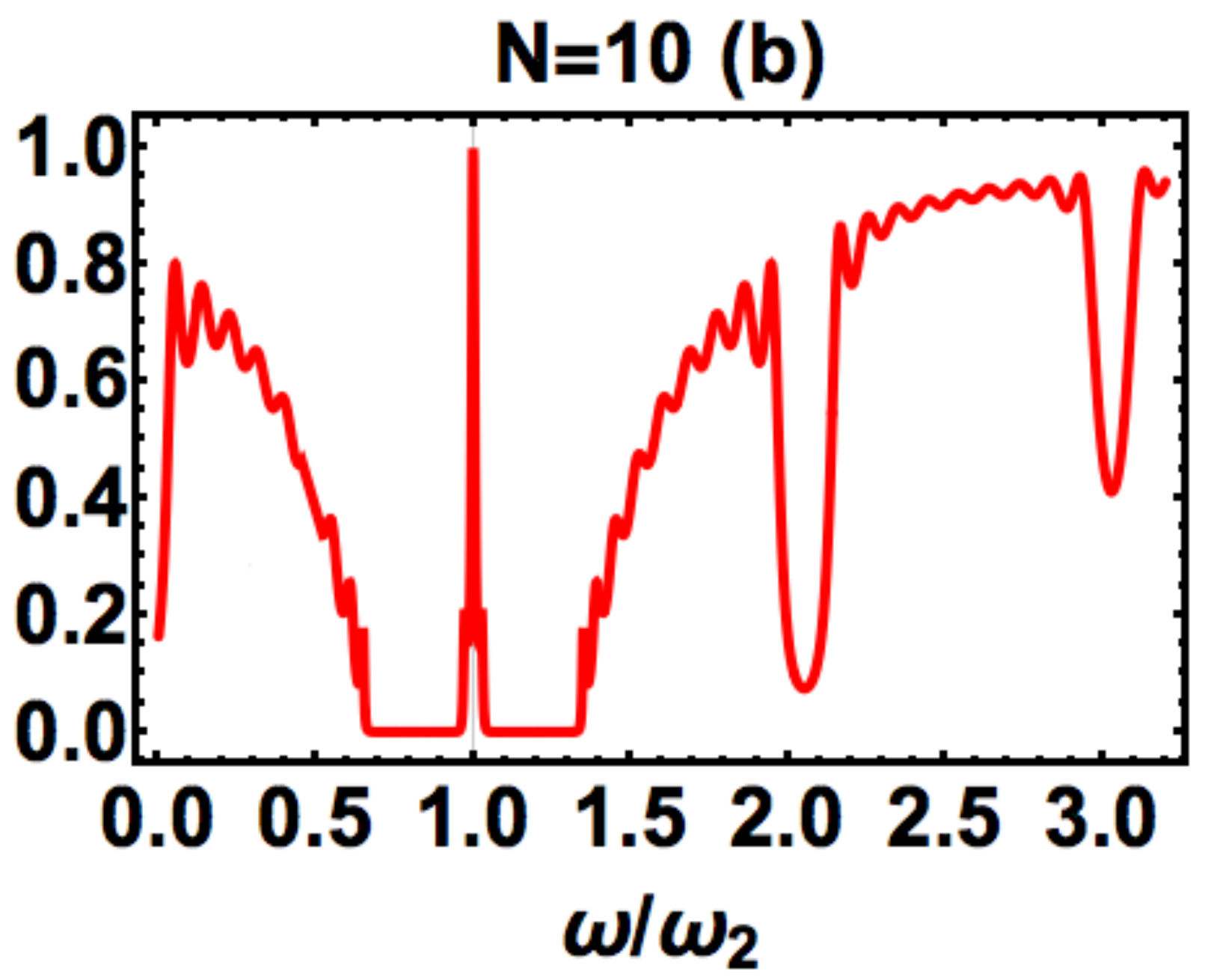}&
   \includegraphics[width=1.6in, height=1.5in]{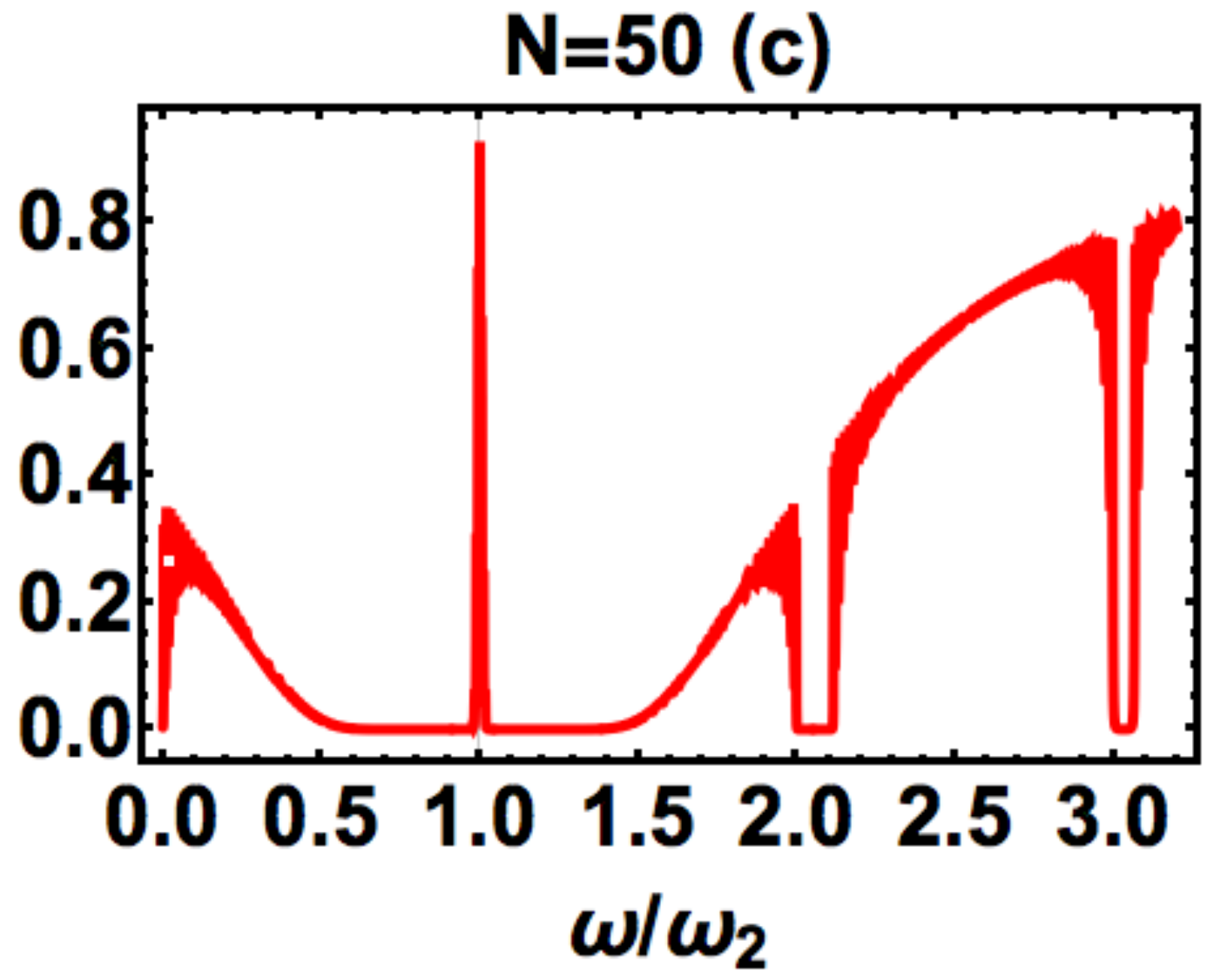}&
      \includegraphics[width=1.5in, height=1.85in]{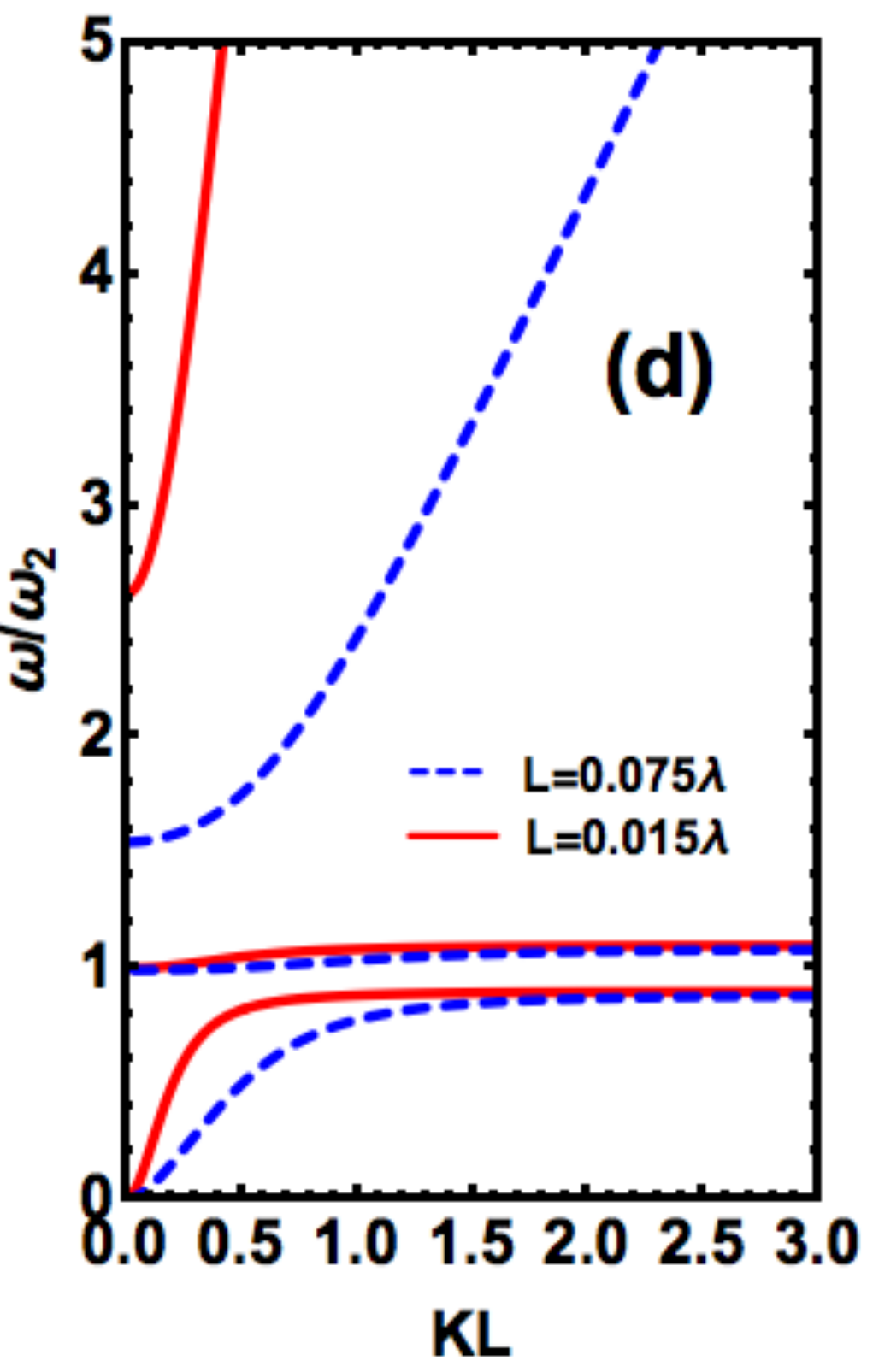} 
  \end{tabular}
\captionsetup{
  format=plain,
  margin=1em,
  justification=raggedright,
  singlelinecheck=false
}
 \caption{(Color online) Single-photon transmission and reflection in a symmetric waveguide. (a) $N=1$ (b) $N=10$ and (c) $N=50$ periodically ordered identical emitters. (d) Dispesion profile for two different inter-atomic separations. Here $\Gamma_{R}=\Gamma_{L}=0.4\omega_{2}$ and the rest of the parameters are same as used in Fig.~4.}
 \label{Fig5}
\end{figure*}
In Fig.~\ref{Fig4}(d) we plot the dispersion relation for two different inter-atomic separations. We note that for large separations the number of dispersion branches  increases. For smaller separations we also see that tiny band gaps are formed. Thus small inter-atomic separations along with small back reflections can create destructive interference sufficient to form forbidden bands for single photon transport. This result agrees with the band structure obtained for the case of two-level atoms when $\Omega=0$~\cite{mirza2017chirality}. The key difference between the band structures for two and three-level emitters, is that for small interatomic separations, two-level atoms form forbidden bands centered at the atomic transition frequency, whereas for three-level atoms a dispersion branch appears at $\omega=\omega_{1}$.

\subsection{Symmetric waveguides}
We now focus on symmetric waveguides with group velocities $v_{R}=v_{L}$ (equivalently $\Gamma_{L}=\Gamma_{R}$). For a single emitter, the transmission and reflection coefficients take the form
\begin{equation}
t=\frac{\delta_{2}(\delta_{2}+i\gamma_{2}/2)-(\Omega/2)^{2}}{\delta_{2}(\delta_{2}+i(\gamma_{2}/2+(\Gamma_{R}+\Gamma_{L})/4))-(\Omega/2)^{2}}, \ r=\frac{-i\delta_{2}\sqrt{\Gamma_{R}\Gamma_{L}}/2}{\delta_{2}(\delta_{2}+i(\gamma_{2}/2+(\Gamma_{R}+\Gamma_{L})/4))-(\Omega/2)^{2}}.
\end{equation}
In Fig.~\ref{Fig5} we present plots of the transmission. In the single emitter case, the transmission coefficient exhibits the standard EIT pattern with complete transmission at resonance. As the emitter number is increased, a band structure emerges. Comparing this band structure with the small back reflection case, we find that the separation between the peaks of suppressed transmission is larger than for the symmetric waveguide problem.  For symmetric waveguides ($q_{L}=q_{R}=q$) the dispersion relation takes the form 
\begin{equation}
{\rm cos}(KL)={\rm cos}(qL)+\left(\frac{\delta_{2}\Gamma}{2\Lambda^{2}}\right){\sin}(qL)
\end{equation}
In Fig.~\ref{Fig5}(d) we plot the dispersion relation.
Similar to the small back reflection case, we conclude that inter-atomic separation can be used as a probe to engineer the dispersion properties of single photons. However, the general features of the band structure are considerably changed. In particular, the dispersion branches in the small back reflection case are  removed. Additionally, for large back reflections and both inter-atomic separations, the width of the forbidden bands is increased for symmetric waveguides.

\section{Disorder}
\subsection{Evidence for localization}
For chiral waveguides, we calculated the localization length analytically and were able to establish the existence of localization. However, for bidirectional waveguides, an analogous analysis is not straightforward. Instead, we seek  evidence for localization by numerically demonstrating that (\ref{llc}) holds. To this end, we plot $\langle \ln T\rangle$ as a function of the number of atoms $N$  in Fig.~\ref{Fig6}. Four cases of interest have been considered: position disorder in symmetric waveguides or with small back reflections, and frequency disorder in symmetric waveguides or with small back reflections. We see that $\langle \ln T\rangle$ decreases linearly as a function of $N$, consistent with (\ref{llc}). Based on this result and using (\ref{llc}) we compute the localization length.

\begin{figure}[t]
\centering
  \begin{tabular}{@{}cccc@{}}
\includegraphics[width=2.25in, height=1.5in]{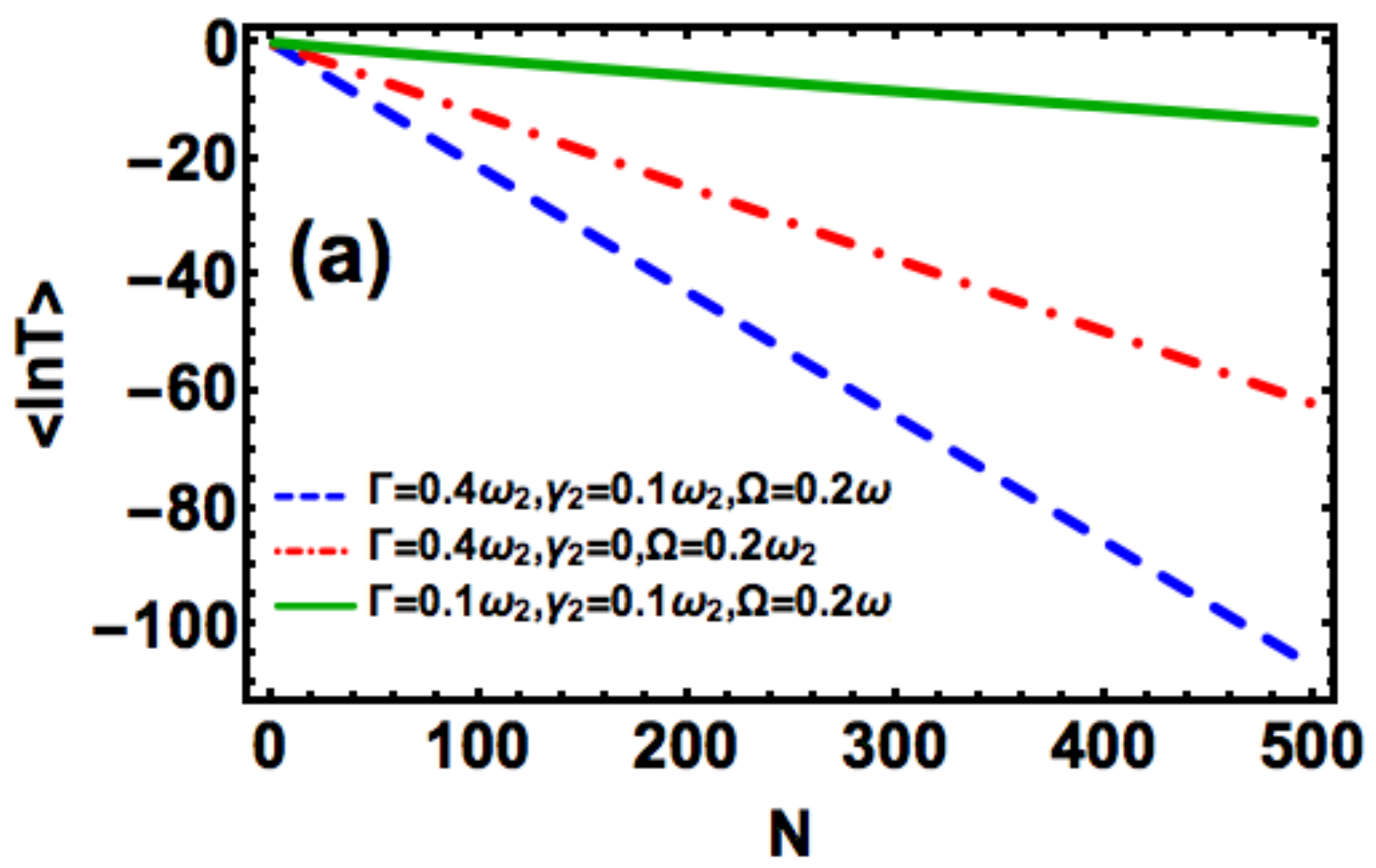} 
  \hspace{-1mm}\includegraphics[width=2.17in, height=1.48in]{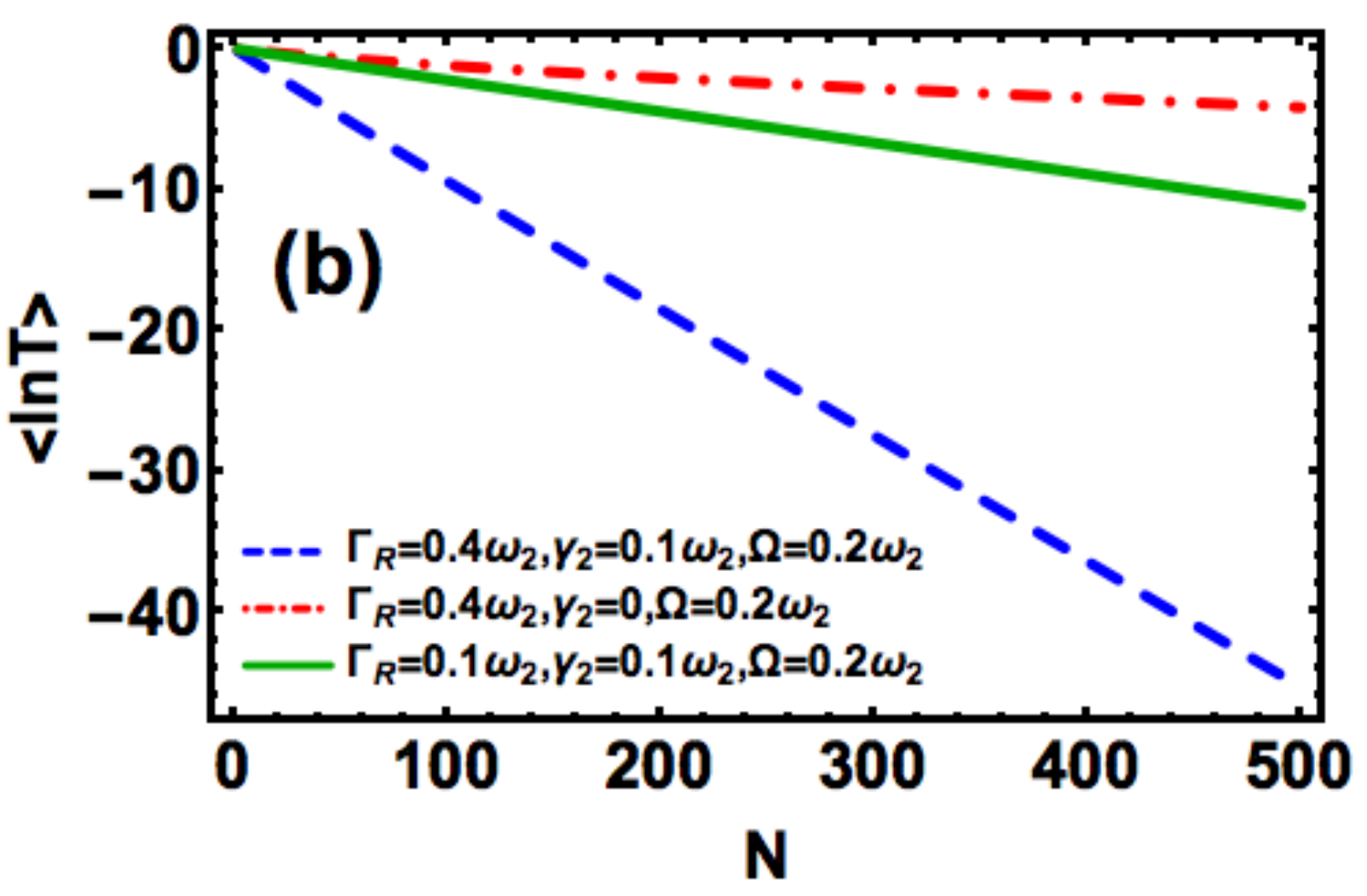}
   \hspace{-1mm}\includegraphics[width=2.15in, height=1.5in]{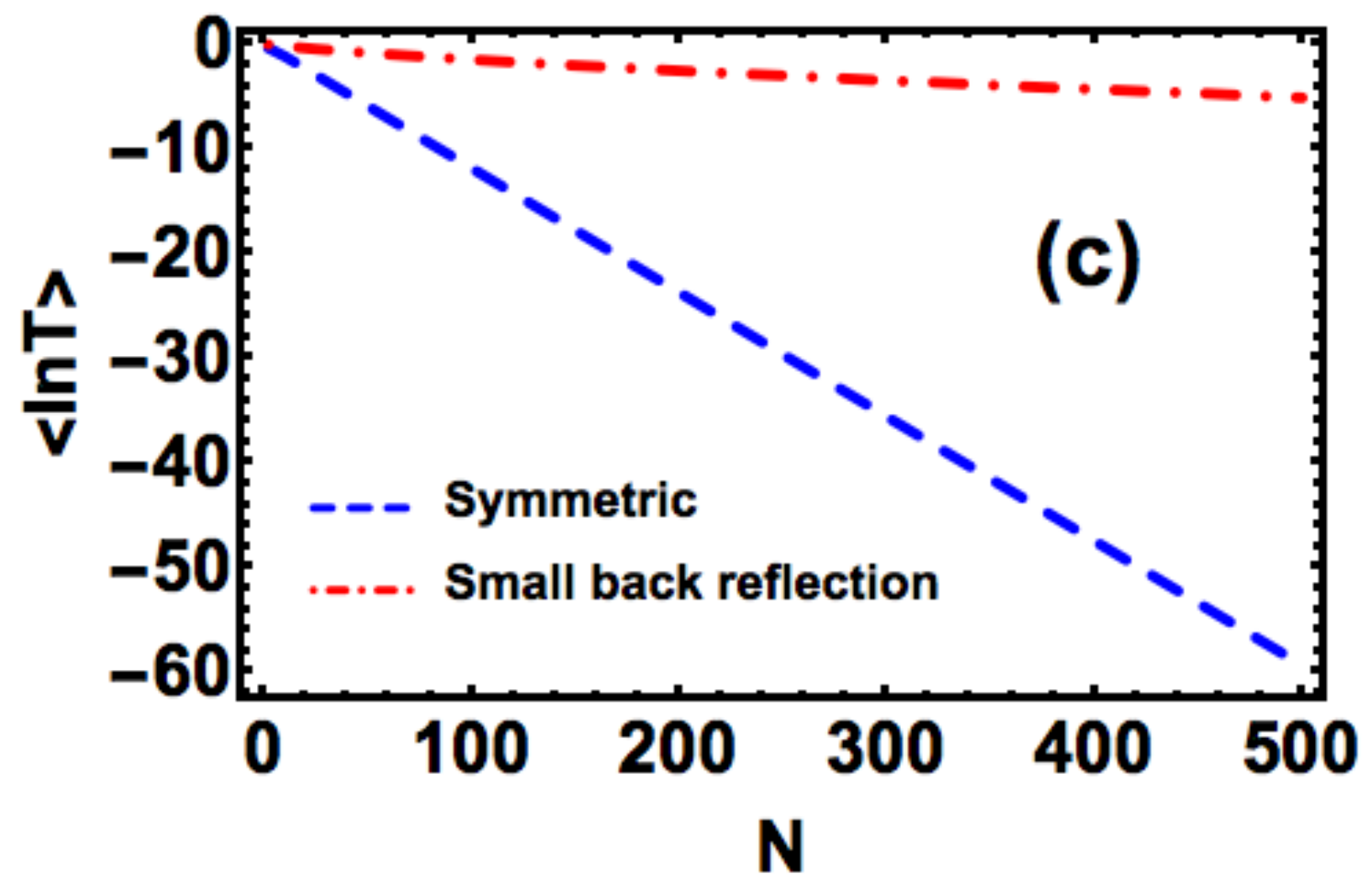} 
  \end{tabular}
\captionsetup{
  format=plain,
  margin=1em,
  justification=raggedright,
  singlelinecheck=false
}
\caption{(Color online) 
Dependence of $\langle \ln T \rangle$ on the number of atoms $N$. (a) and (b) correspond to position disorder with $\omega=1.5\omega_{2}$, mean interatomic separation $\lambda/2$ and strength of the disorder $\sigma=\lambda$. (a) Symmetric waveguide and (b) Small back reflections with $\Gamma_{L}=0.1\Gamma_{R}$.
(c) Frequency disorder. A periodic atomic array is considered with a lattice constant $L=\lambda/2$. The mean disorder for small back reflections (symmetric waveguides) is $3\Gamma_{R}$ ($3\Gamma$) and the strength of the disorder is $\sigma=\Gamma_{R}$ ($\Gamma$). 
We have set $\gamma_{2}=0$ (no spontaneous decay). In all plots, {we have performed the average over $10^{5}$} realizations of the disorder; the error bars are too small to be shown.}
\label{Fig6}
\end{figure}
\subsection{Small back reflections}
\subsubsection{position disorder}
 We begin with the case of position disorder and consider in Fig.~\ref{Fig7}(a) the transmission  as a function of of incoming photon frequency for a ten atom chain. We find that in the presence of position disorder the EIT peak survives but the pattern of small transmission peaks superimposed on EIT, observed in the periodic situation, vanishes. In Fig.~\ref{Fig7}(b) we plot the localization length $\xi$ as a function of $\omega$. We observe $\xi$ takes a very large value at resonance when the system is completely transmitting. Following a pattern similar to the transmission plot (Fig.~\ref{Fig7}(a)), we see that at $\omega\simeq 1.1\omega_{2}$ and $0.9\omega_{2}$, localized photonic states are formed. Away from these points, the system shows enhanced transmission and $\xi$ grows. In Fig.\ref{Fig7}(c) we plot the dependence of $\xi$ on the disorder strength $\sigma$ for a $10^{3}$ emitter chain. Assuming a detuned system, we take into consideration the cases of strong and weak atom-waveguide couplings for a fixed Rabi frequency. In all cases, we find that the localization length is initially large and then decreases for small disorder strengths. For larger disorder, $\xi$ takes on a small and an almost constant value. Additionally, we point out that $\xi$ is smaller for strong as compared to weak atom-waveguide couplings. We also investigate the influence of spontaneous emission (non-zero $\gamma_{2}$) on $\xi$. We find that the presence of spontaneous emission reduces the localization length noticeably and as a result, the dependence of $\xi$ on $\sigma$ becomes quite weak.
\begin{figure*}
\centering
  \begin{tabular}{@{}cccc@{}}
   \includegraphics[width=2.75in, height=2in]{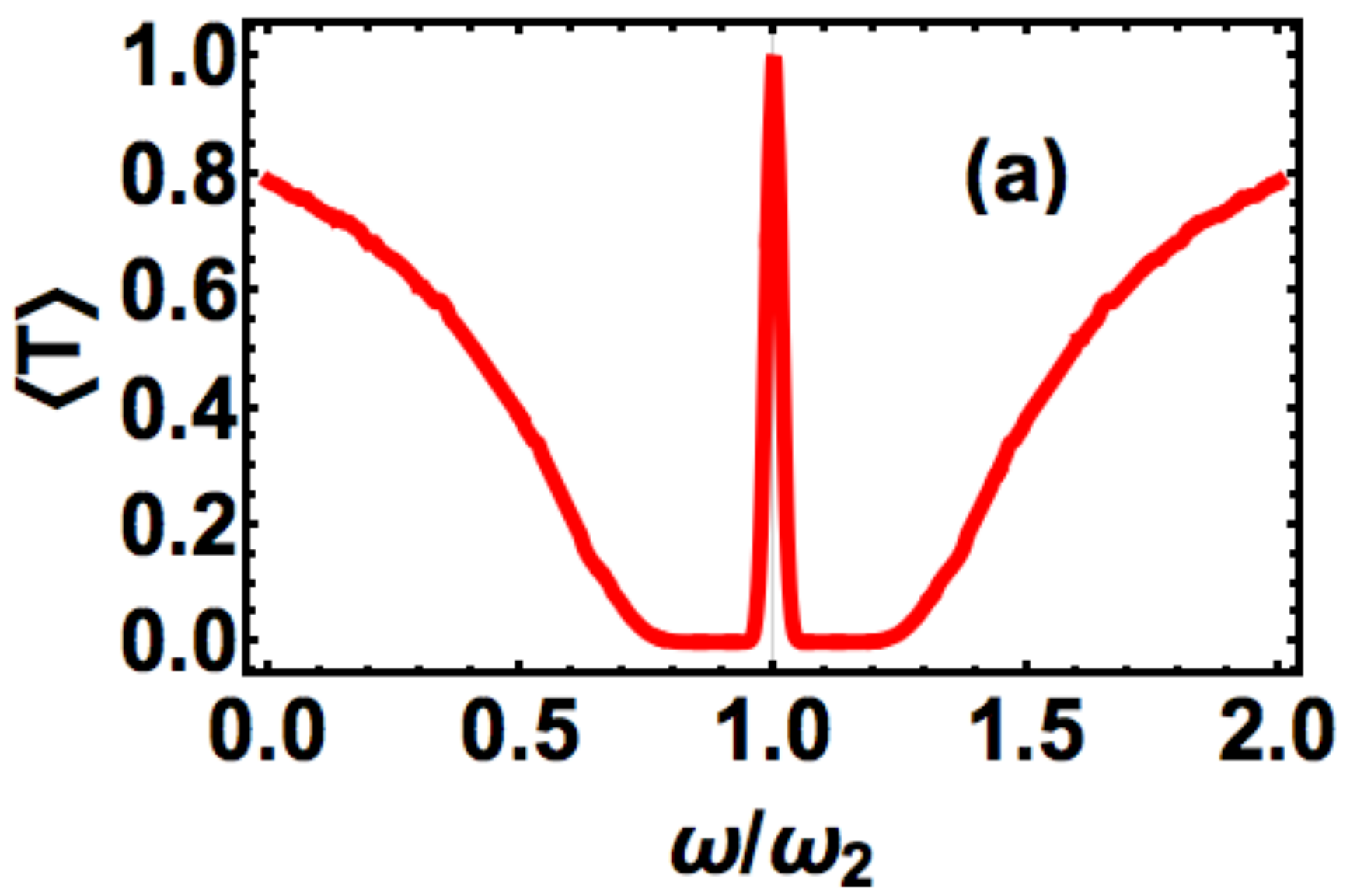} 
  \includegraphics[width=2.75in, height=2in]{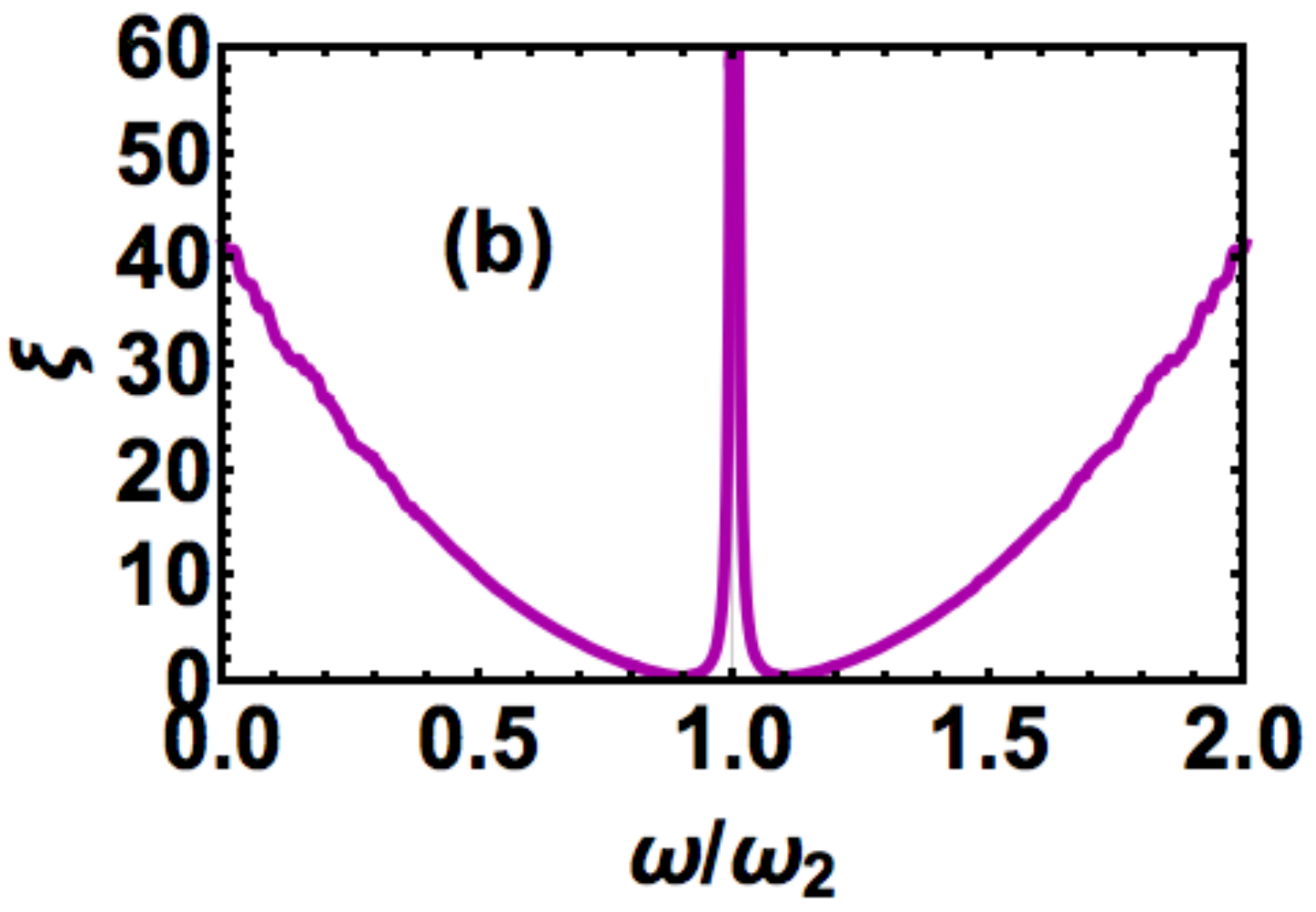}\\
   \includegraphics[width=2.75in, height=2in]{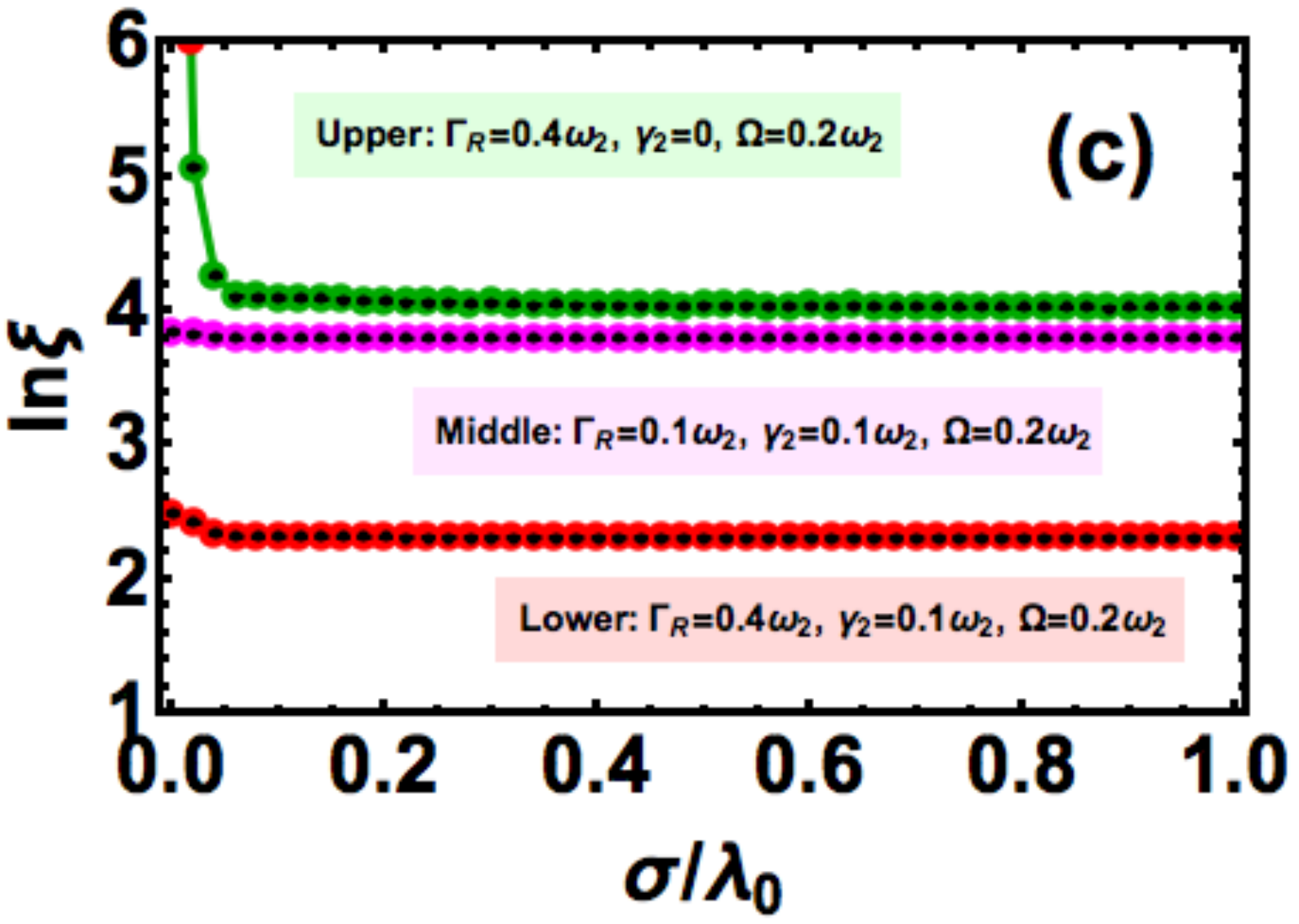}
   \includegraphics[width=2.75in, height=2in]{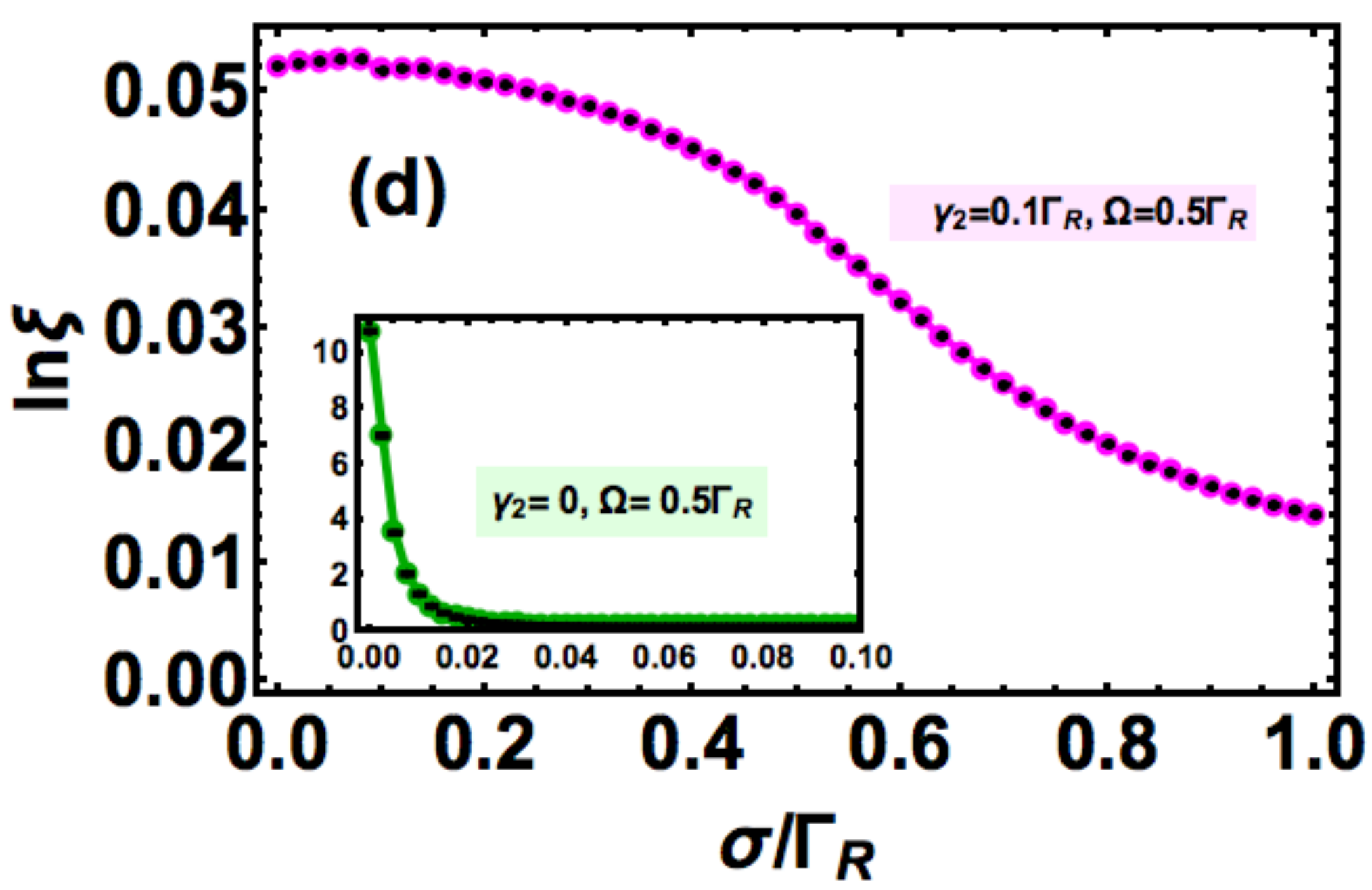}
  \end{tabular}
\captionsetup{
  format=plain,
  margin=1em,
  justification=raggedright,
  singlelinecheck=false
}
 \caption{(Color online) Single-photon transmission and localization length for a waveguide with small back reflections. Here  $v_{L}=10v_{R}$ (or $\Gamma_{L}=0.1\Gamma_{R}$). For (a) and (b), $N=10$, $\gamma_{2}=0$, the mean spacing is $\lambda/2$ and the strength of the disorder is $2\lambda$. (a) Average transmission $\langle T\rangle$ and (b) localization length $\xi$ for position disorder. Averages for these two plots have been performed over 500 realizations. (c) Localization length for position disorder as a function of disorder strength with mean spacing $\lambda/2$ and $\omega=1.5\omega_{2}$. (d) $\xi$ versus $\sigma$ for frequency disorder with lattice constant $\lambda/2$ and mean frequency is $3\Gamma_{R}$. In plots (c) and (d) $N=10^{3}$ and average is carried out on $10^{4}$ realizations.  }\label{Fig7}
\end{figure*}

\subsubsection{Frequency disorder}
In Fig.~\ref{Fig7}(d) we plot $\xi$ as a function of $\sigma$ for frequency disorder. Here we consider the mean frequency to be detuned. For $\gamma_{2}=0$, $\xi$ decreases rapidly and nearly vanishes as $\sigma$ is slightly increased (see plot inset). For $\gamma_{2}\neq 0$ the transmission takes very small values for all values of $\sigma$,  but decays as $\sigma$ increases.

\subsection{Symmetric waveguides}
\begin{figure*}
\centering
  \begin{tabular}{@{}cccc@{}}
   \includegraphics[width=2.75in, height=2in]{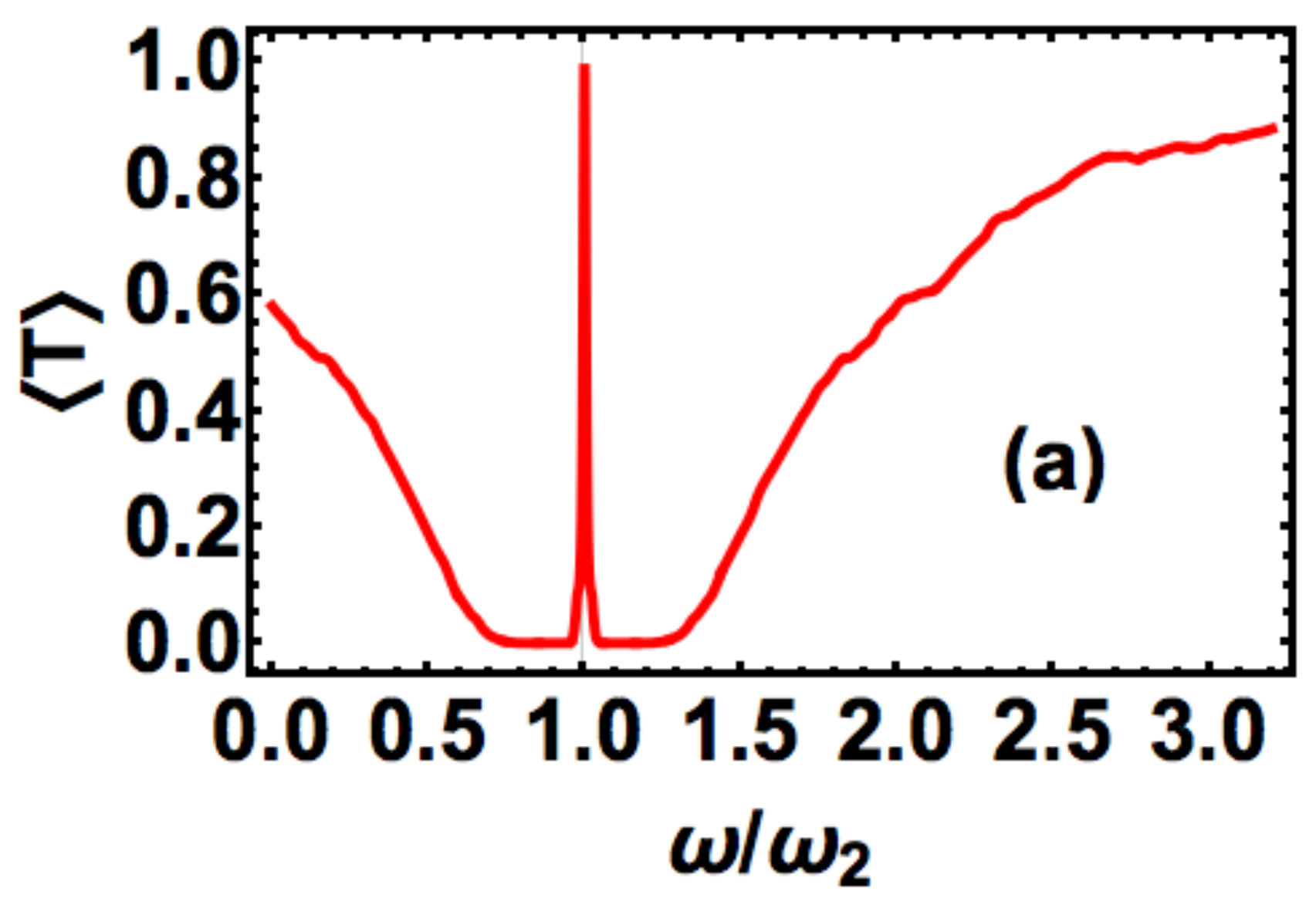} 
  \includegraphics[width=2.75in, height=2in]{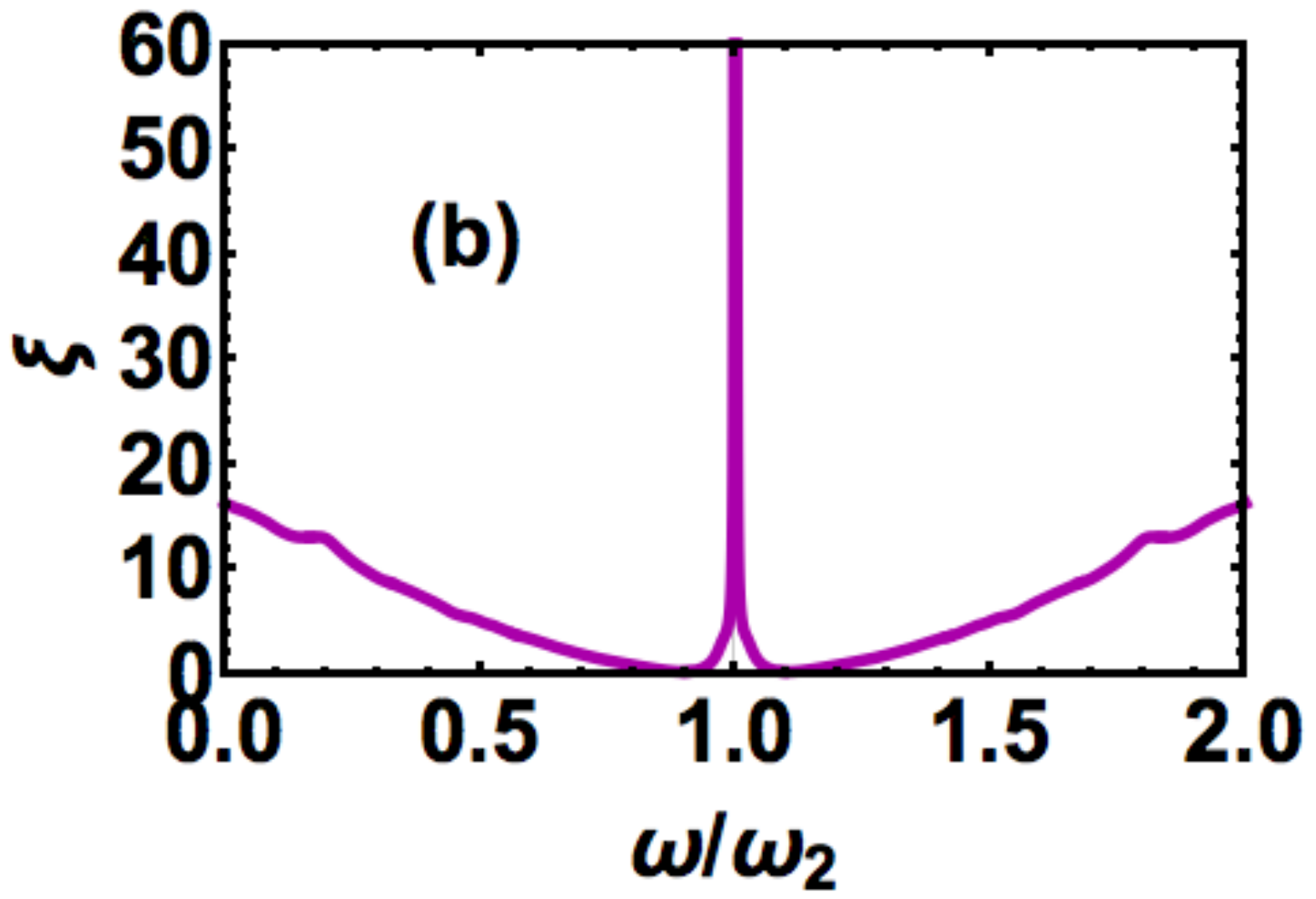}\\
   \includegraphics[width=2.75in, height=2in]{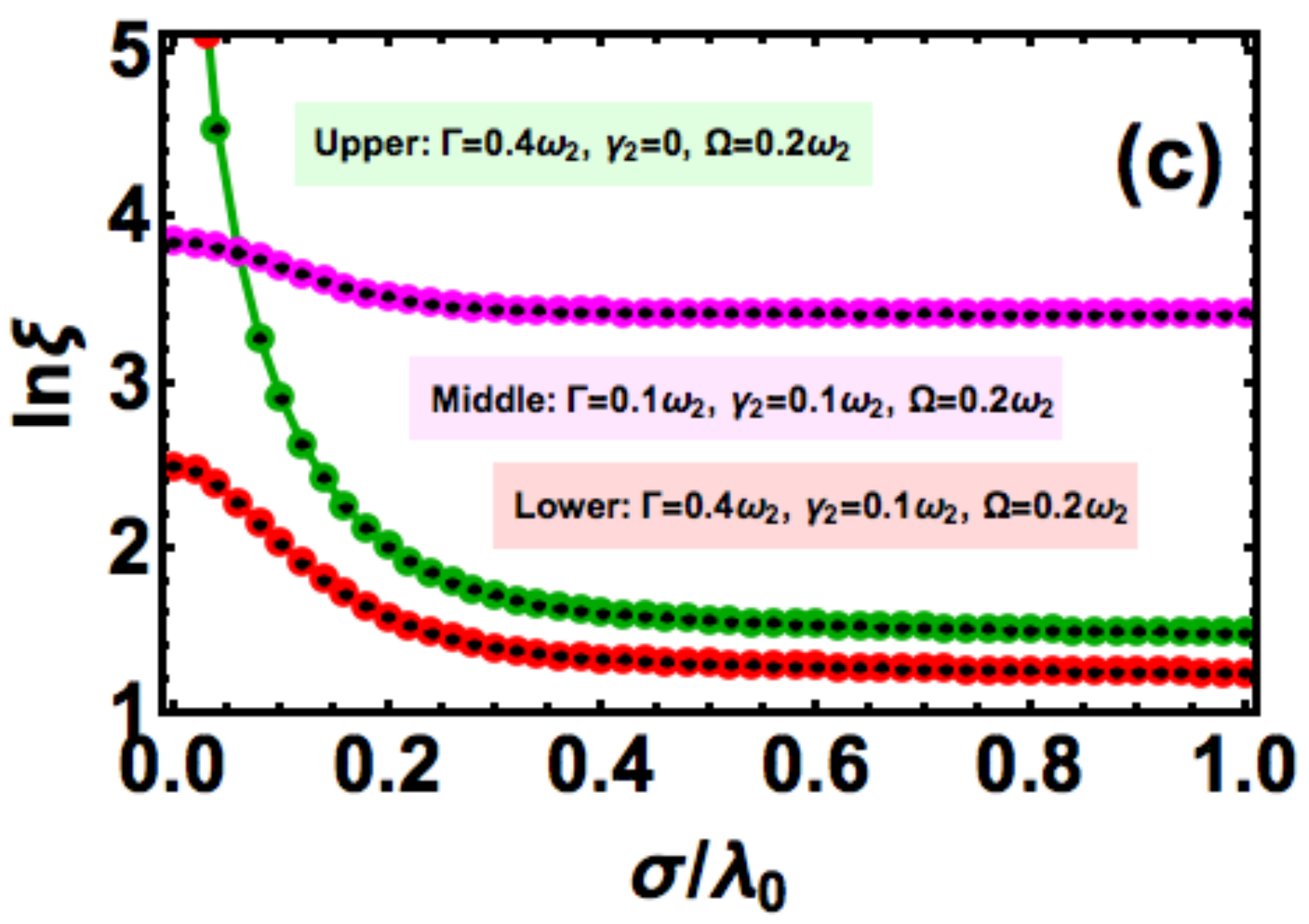}
   \includegraphics[width=2.75in, height=2in]{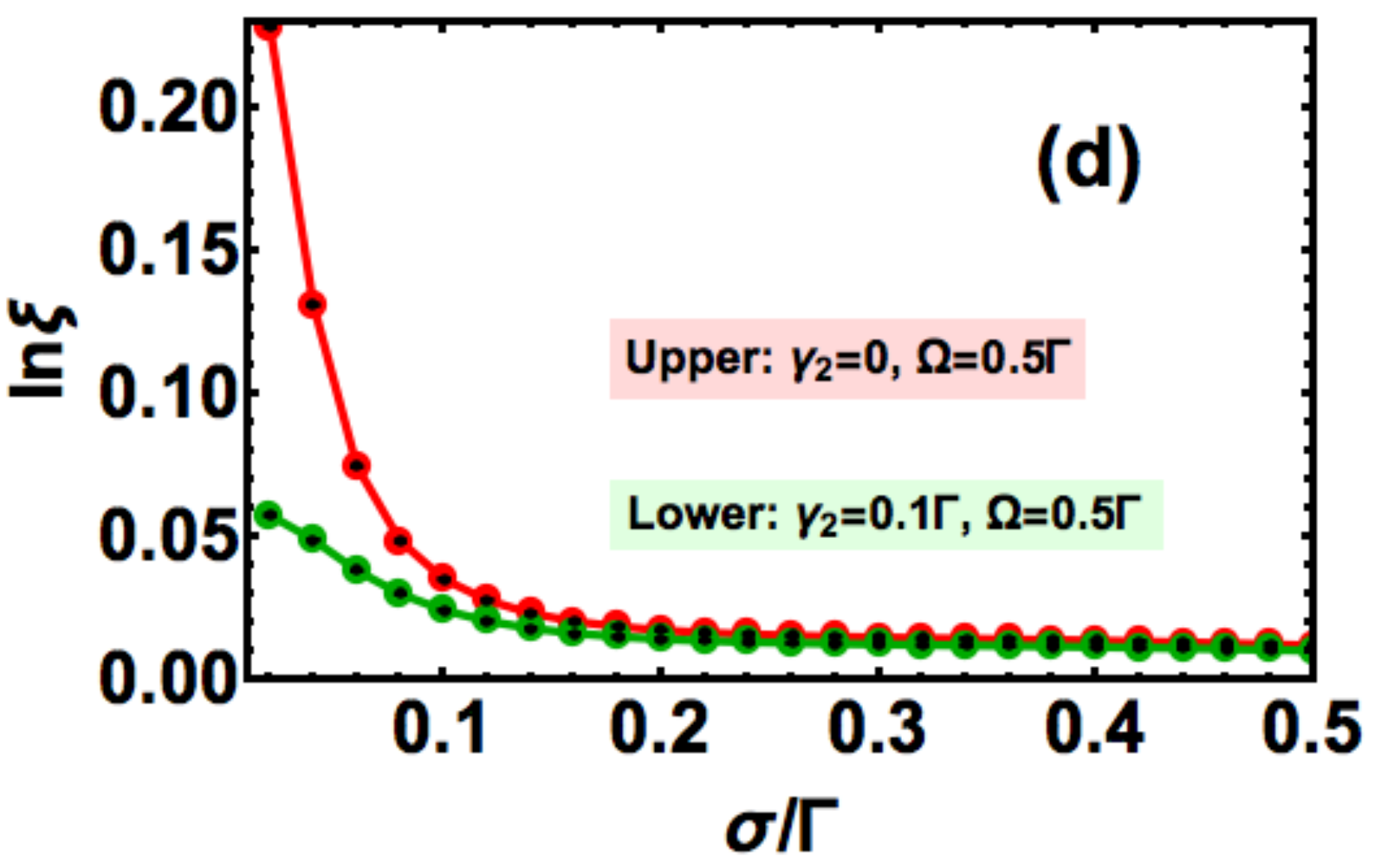}
  \end{tabular}
\captionsetup{
  format=plain,
  margin=1em,
  justification=raggedright,
  singlelinecheck=false
}
 \caption{(Color online) Localization in a symmetric waveguide.  The parameters are the same as in Fig.~\ref{Fig7}.}\label{Fig8}
\end{figure*}

Following the discussion of waveguides with small back reflections, we now focus on the case of symmetric waveguides. Our results  are presented in Fig.~\ref{Fig8}. We  note that the behavior of the transmission and localization length is similar to the case of small back reflections. However, for symmetric waveguides, the scale of $\xi$ is decreased. Finally, we mention that the behavior of the localization length can become increasing as a function of $\sigma$ if we chose the parameters so that $\omega$ lies in a band gap of the corresponding periodic problem. 

\section{Discussion}
We have investigated single photon transport in chiral and bidirectional disordered waveguide QED architectures with $\Lambda$-configured three-level emitters. In particular, we analyzed the band structure for  periodically spaced atoms and considered the effects of disordered atomic positions and atomic transition frequencies. In the periodic chiral case, the system exhibited EIT but failed to exhibit a band structure. We found that transport is immune to position disorder in  chiral waveguides, but displays localization for frequency disorder.

For waveguides with small back reflections and with periodic atomic arrangements, we observed the existence of multiple resonances superimposed on the EIT pattern. For the case of symmetric waveguides, the resonances lead to relatively small bandgaps. In this setting, the dispersion relation shows sensitive dependence on the interatomic separation. Next, we found that both position and frequency disorder can localize single photons. The localization length is smallest at the two frequencies near the EIT peak. Moreover, our results for the transmission and localization length exhibit suppression of photon transport. This suppression is enhanced for strong emitter-waveguide coupling. Finally, we found that the presence of spontaneous emission reduces transmission considerably. In comparison to the case of two-level atoms \cite{mirza2017chirality}, we find that  the details of the dependence of the transmission and localization length on the frequency $\omega$ are different. For instance, in the case of position-disordered two-level atoms in bidirectional waveguides, a null-transmission band is formed at the atomic resonant frequency. In addition, for a periodic chain of two-level atoms with small inter-atomic separations, a forbidden band at the system resonance is formed.

\acknowledgments

This work was supported in part by the NSF grants DMR-1120923 and DMS-1619907.

\bibliographystyle{ieeetr}
\bibliography{Paper}
\end{document}